\documentclass[10pt, journal, compsoc]{IEEEtran}

\usepackage{mystyle}

\togglefalse{complete-version}
\togglefalse{highlight-revision}

\def\doctitle{On-the-Fly Syntax Highlighting:\\ Generalisation and Speed-ups}

\def\docauthors{Marco Edoardo Palma, Pasquale Salza, Harald C. Gall}

\def\dockeywords{%
Syntax highlighting, neural networks, deep learning, regular expressions
}

\StrSubstitute{\doctitle}{\\}{ }[\cleandoctitle]
\StrSubstitute{\dockeywords}{.}{}[\cleandockeywords]
\hypersetup{
	pdftitle={\cleandoctitle},
	pdfauthor={\docauthors},
	pdfkeywords={\cleandockeywords}
}

\DeclareAcronym{api}{
	short = API,
	long = {Application Program Interface}
}

\DeclareAcronym{bert}{
	short = BERT,
	long = {Bidirectional Encoder Representations from Transformers}
}

\DeclareAcronym{ast}{
	short = AST,
	long = {Abstract Syntax Tree}
}

\DeclareAcronym{cfg}{
	short = CFG,
	long = {Control Flow Graph}
}

\DeclareAcronym{gpt}{
	short = GPT,
	long = {Generative Pretrained Transformer}
}

\DeclareAcronym{ir}{
	short = IR,
	long = {Information Retrieval}
}

\DeclareAcronym{lstm}{
	short = LSTM,
	long = {Long Short-Term Memory}
}

\DeclareAcronym{nn}{
	short = NN,
	long = {Neural Network}
}

\DeclareAcronym{rnn}{
	short = RNN,
	long = {Recurrent Neural Network}
}

\DeclareAcronym{brnn}{
	short = BRNN,
	long = {Bidirectional Recurrent Neural Network}
}

\DeclareAcronym{cnn}{
	short = CNN,
	long = {Convolutional Neural Network}
}

\DeclareAcronym{tf-idf}{
	short = tf-idf,
	long = {term frequency–-inverse document frequency}
}

\DeclareAcronym{anova}{
	short = ANOVA,
	long = {ANalysis Of VAriance}
}

\DeclareAcronym{ta}{
	short = TA,
	long = {Task-Adaptive}
}

\DeclareAcronym{gru}{
	short = GRU,
	long = {Gated Recurrent Unit}
}

\DeclareAcronym{sh}{
	short = SH,
	long = {syntax highlighting}
}

\DeclareAcronym{re}{
	short = regex,
	short-plural = es,
	long = {regular expression}
}

\DeclareAcronym{om}{
	short = OM,
	long = {Oracle Methods}
}

\DeclareAcronym{eta}{
	short = ETA,
	long = {Extended Token Annotation}
}

\DeclareAcronym{heta}{
	short = HETA,
	long = {Highlighted Extended Token Annotation}
}

\DeclareAcronym{ide}{
	short = IDE,
	long = {Integrated Development Environment}
}

\DeclareAcronym{bf}{
	short = BF,
	long = {brute-force}
}

\DeclareAcronym{peg}{
	short = PEG,
	long = {Parsing Expression Grammar}
}

\DeclareAcronym{dl}{
	short = DL,
	long = {deep learning}
}

\begin{document}

\title{\doctitle}

\author{
	Marco Edoardo Palma, Alex Wolf, Pasquale Salza, and Harald C. Gall

	\IEEEcompsocitemizethanks{ %
		\IEEEcompsocthanksitem The authors are with the University of Zurich, Zurich, Switzerland. E-mail: \href{mailto:marcoepalma@ifi.uzh.ch}{marcoepalma@ifi.uzh.ch}, \href{mailto:wolf@ifi.uzh.ch}{wolf@ifi.uzh.ch}, \href{mailto:salza@ifi.uzh.ch}{salza@ifi.uzh.ch}, \href{mailto:gall@ifi.uzh.ch}{gall@ifi.uzh.ch}.
	}
}

\IEEEtitleabstractindextext{
    \begin{abstract}
On-the-fly syntax highlighting is the task of rapidly associating visual secondary notation values with each character of a language derivation.
Research in this domain is driven by the prevalence of online software development tools, which frequently display source code on screen and heavily rely on syntax highlighting mechanisms.
In this context, three contrasting demands confront resolvers in this space: speed, accuracy, and development costs.
Speed constraints are essential to ensure tool usability, manifesting as responsiveness for end users accessing online source code and minimising system overhead.
Simultaneously, achieving precise highlighting is critical for enhancing code comprehensibility.
Nevertheless, obtaining accurate results necessitates the capacity to perform grammatical analysis on the code under consideration, even in cases of varying grammatical correctness.
Furthermore, addressing the development costs of such resolvers is imperative, given the multitude of programming language versions.
The current \emph{state-of-the-art} approach in this field leverages the original lexer and parser of programming languages to create syntax highlighting oracles, subsequently used for training base Recurrent Neural Network models.
As the question of the generalisation of such a solution persists, this paper addresses this aspect by extending the original work to three additional mainstream programming languages and conducting a comprehensive review of the outcomes.
Moreover, the original limitations in evaluation performance and training costs are mitigated through the introduction of a novel Convolutional based Neural Network model.
Notably, this paper explores an area where previous research has remained silent: the performance advantages of running such models on GPUs.
The evaluation identifies the new CNN based implementation as being significantly faster than the \emph{state-of-the-art} approaches, whilst delivering the same near-perfect level of accuracy.
\end{abstract}
    \begin{IEEEkeywords}
    \dockeywords
\end{IEEEkeywords}

}
\maketitle

\section{Introduction}
\label{sec:introduction}
\Ac{sh} is the practice of visually annotating code by associating distinct colours with specific language sub-productions, enhancing code comprehensibility~\cite{sarkar_impact_2015}. Code is presented in a multitude of online contexts, such as code review workflows, repository file browsers, and various forms of code snippets, all of which benefit from \ac{sh} mechanisms~\cite{10.1145/3540250.3549109}. Importantly, these platforms employ \ac{sh} \enquote{On-the-Fly}, meaning that \ac{sh} resolvers compute the highlighting for code just before it's displayed to the user.

The design choice of dynamic \ac{sh}, driven by space constraints and the inability to cache notations, places \ac{sh} resolvers in a challenging position. These resolvers must operate efficiently, responding swiftly to a high frequency of requests, thus enhancing the platform's usability. Additionally, they are expected to deliver accurate highlighting, ensuring that code sub-productions are correctly bound to their respective \ac{sh} classes or colours. Achieving this level of accuracy requires a grammatical analysis of the code. However, a full parsing process on the input file is infeasible in this context, given the time constraints and the likelihood of incorrect derivations~\cite{10.1145/3540250.3549109}. Moreover, the development costs associated with these solutions are significant, given the ever-expanding landscape of mainstream programming languages and their versions.

Traditionally, developers relied on manual creation of complex systems of regular expressions to achieve \ac{sh}, despite its tedium and inaccuracy. While effective, this approach gave way to more recent efforts that reduce development costs, improve \ac{sh} accuracy, and expand coverage of grammatical sub-productions. The contemporary \emph{state-of-the-art} approach~\cite{10.1145/3540250.3549109} treats \ac{sh} as a machine learning translation problem. It leverages the language's original lexer to tokenize code, subsequently binding each token to a \ac{sh} class. Development costs are reduced by utilising the language's original parser and lexer for creating a \ac{sh} oracle, requiring developers only to design a deterministic \ac{ast} walker for each language. This approach excels not only in accuracy and coverage but also in handling incorrect input sequences. In cases of incorrect derivations, it makes statistically relevant \ac{sh} inferences for final colouring, and the model's efficiency is determined by the machine learning model's performance.

While the \emph{state-of-the-art} approach has been validated for three mainstream programming languages, namely \emph{Java}, \emph{Kotlin}, and \emph{Python3}~\cite{10.1145/3540250.3549109}, its generalisation across a broader spectrum of programming languages remains unconfirmed. This paper seeks to address this question by developing formal \ac{sh} models for \emph{JavaScript}, \emph{C++}, and \emph{C\#} and comparing their performances to the \emph{state-of-the-art} approach.

The original approach~\cite{10.1145/3540250.3549109} suggested the use of baseline \ac{rnn} models, with bidirectional variants \ac{brnn} achieving the best performance. However, the pursuit of even more efficient models remained an open question. Therefore, this paper introduces a faster \ac{cnn} based model and evaluates its performance across all tested languages.

The paper's main findings confirm the generalisation of the original approach on the newly evaluated languages and demonstrate how the proposed \ac{cnn} approach maintains accuracy while significantly accelerating computations.

Furthermore, this paper addresses a previously unexplored aspect: the performance advantages of running these models on \emph{GPU}s, shedding light on the efficiency gains achieved by leveraging high-performance hardware.
Evidence shows how \emph{GPU} evaluation can lead to significant gains in the prediction speed of such statistical resolvers, especially in the case of the novel \ac{cnn} implementation.

In summary, the primary contributions of this paper are:
\smallskip
\noindent
\begin{itemize}

    \item A dataset for \ac{sh} benchmarking that has been extended to include \emph{JavaScript}, \emph{C++}, and \emph{C\#}, building upon the previous collection of \emph{Java}, \emph{Kotlin}, and \emph{Python}.
    \item An analysis of the generalisation of the \emph{state-of-the-art} \ac{sh} approach to the extended dataset.
    \item The introduction of a more efficient \ac{cnn} prediction model for \ac{sh}.
    \item A comprehensive comparison of the newly developed \ac{cnn} model with the \emph{state-of-the-art} \ac{rnn} and \ac{brnn} approach in terms of accuracy, coverage, and execution time.
    \item An in-depth performance analysis of both approaches in handling incorrect or incomplete language derivations.
    \item An evaluation of the speed advantages gained by running these models on GPUs, highlighting the efficiency improvements enabled by high-performance hardware.
\end{itemize}
\smallskip


The implementation, benchmark datasets, and results are available in the replication package of this work~\cite{replicationpackage}.

This paper is organized in the following manner: Section 2 outlines the approach's design. Section 3 details the experimental framework, while Section 4 provides and analyses the findings. Section 5 reviews work related to this study, and Section 6 offers a conclusion that summarises the key outcomes and contributions, and considers directions for future investigations in this field.

\section{Approach}
\label{sec:approach}


The approach outlined in this paper is dedicated to addressing the challenges presented in this field. The primary goal is to expand the dataset following the strategy introduced in the \emph{state-of-the-art} approach~\cite{10.1145/3540250.3549109} and introduce a novel, \ac{cnn}-based, approach to enhance the evaluation speed.

The devised strategy continues to focus on the development of \acp{cnn} models capable of statistically inferring the optimal behaviour of \ac{bf} models. To achieve this objective, an oracle of \acf{sh} solutions is created by employing the language-specific \ac{bf} resolver.

The subsequent sections provide a comprehensive specification of both the \ac{bf} and \ac{cnn} models, along with the rationale behind their design.

\subsection{Creation of Syntax Highlighting Oracles}
\label{sec:approach:bf}
In this study, language highlighting oracles are generated for three new programming languages: \emph{JavaScript}, \emph{C++}, and \emph{C\#}, extending from the original dataset consisting of \emph{Java}, \emph{Kotlin}, and \emph{Python}. This is achieved through the application of the \ac{bf} model, which is tailored for each specific language to compute precise \ac{sh} assignments for individual source code files.

The \ac{bf} model involves two key components. Initially, it utilises the existing lexer of the target programming language to tokenize the source code, resulting in a stream of tokens. Afterwards, these tokens are structured into an \ac{ast} using the language's parser.

Subsequently, a tree-walking process harnesses the structural information within the \ac{ast} to assign each token to its corresponding \ac{sh} class based on the grammatical context in which it is employed. The primary objective is to ensure accurate association between each token and its designated \ac{sh} class, thereby achieving the highest possible highlighting accuracy for a given language~\cite{10.1145/3540250.3549109}.

It is essential to note that the \ac{bf} model requires the implementation of a walker which typically consists of a small number of detection rules, as outlined in the replication package~\cite{replicationpackage}. This approach ensures the deterministic creation of \ac{bf} highlighters and demands only a fundamental understanding of the language's grammar, as the core lexer and parser tools specific to each language are reused. This method represents a significant departure from the conventional and error-prone processes of defining complex systems of regular expressions.

The \ac{bf} model plays a pivotal role in generating the \emph{oracle}, which is a compilation of the source code files for the target programming language, along with their corresponding \ac{sh} assignments. To accomplish this, each source code file undergoes two primary steps.

\textbf{Tokenisation}: The file is processed through the language's lexer, which dissects the code into tokens. Each token is then transformed into an \acf{eta} entity, represented as $\textit{\acs{eta}} = \{i_s, i_e, t, tr\}$. In this representation:
\begin{inparaenum}
    \item $i_s$ and $i_e$ denote the token's character start and end indexes within the file.
    \item $t$ represents the precise text referenced by the token.
    \item $tr$ signifies the token's unique \emph{Token Rule} encoded as a natural number (or ID) consistently assigned by the language's lexer. This ID signifies the token type, such as keywords, operators, or literals.
\end{inparaenum}
For example, \code{String lang = "Java";} would result in a set of \acp{eta}: \code{\{0, 5, String, 102 \}}, \code{\{7, 10, lang, 102\}}, \code{\{12, 12, =, 73\}}, \code{\{14, 20, "Java", 55\}}, and \code{\{21, 21, ;, 63\}}, where \emph{tr} values represent token types.

\textbf{\ac{ast} Construction}: The language's parser is then employed to structure these tokens into an \ac{ast} based on the language's grammar specifications.

The \ac{bf} resolver function for a given language $L$ can be represented as follows:
\[
{bf}_L: \{c\}, {le}_L, l_L, p_L, {ws}_L \rightarrow \{\textit{\acs{heta}}\}
\]
Where:
\begin{inparaenum}
    \item ${c}$ is the character set of the input file
    \item ${le}_L$ signifies the lexer encoder, which transforms character sets and lexer information into \acp{eta}.
    \item $l_L$ is the lexer of $L$
    \item $p_L$ represents the parser responsible for generating the language-specific \ac{ast}.
    \item ${ws}_L$ denotes the walking strategy that maps \acp{eta} to \emph{\ac{heta}} objects. These \ac{heta} objects extend \acp{eta} by including a \emph{Highlighting Class} ($hc$) corresponding to the grammatical \ac{sh} class to which the token belongs.
\end{inparaenum}
Tokens that do not form part of any grammatical construction are assigned to the unique $hc$ \emph{ANY}, representing unhighlighted text.

This methodology enables the generalisation of \ac{sh} patterns based on the sequence of language features and abstracts away the noise introduced by specific token text features, facilitating the parsing of code.

\subsection{Deep Learning for Syntax Highlighting}
\label{sec:approach}
The \emph{state of the art} approach to efficient syntax highlighting (\ac{sh}) involved the use of \acp{rnn} to map sequences of token rules $\{tr\}$ to sequences of \ac{sh} classes $\{hc\}$, mirroring the process performed by \ac{bf} resolvers. This approach reduced the \ac{sh} task to a statistical inference on the expected grammatical structure of the input token sequence.

The rationale for employing \acp{nn} in this task stemmed from the structured nature of programming language files. Programming languages exhibit the following characteristics:
\begin{inparaenum}
    \item They represent character sequences selected from a finite set of terminal symbols $\{tr\}$.
    \item These sequences adhere to an underlying formal grammar, which imposes a pure ordering function.
\end{inparaenum}
\ac{sh} represents a grammar for which there always exists a correct language derivation when a valid derivation of the original grammar exists. In other words, the \ac{sh} grammar, denoted as $hg$, parses sub-productions $s_{hg}$ of the original grammar $g$ sequentially. These sub-productions are sufficient to discriminate a \emph{tr} subsequence for a target highlighting construction. Alternatively, they map every token not consumable by any $s_{hg}$ to a terminal symbol.
With the \ac{nn} approach to \ac{sh}, the burden of producing the \ac{sh} grammar is shifted from developers to \acp{nn}, which infer it from the observed behaviour of a \ac{bf} model. The \ac{sh} task is effectively transformed into a \emph{sequence-to-sequence} translation task~\cite{sutskever_sequence_2014}, converting sequences of $\{tr\}$ into sequences of $\{hc\}$.
To address this problem reduction, the previous approach employed \acp{rnn}~\cite{cho_learning_2014} for learning the bindings between \ac{sh} sequences. \acp{rnn} are well-suited for sequence translation, iterating through the input sequence to produce a translation unit while retaining information to aid predictions for subsequent inputs.
For grammars producing sequence where binding an \emph{hc} to a \emph{tr} may require looking ahead over an arbitrary number of tokens, the approach turned to \acp{brnn}~\cite{schuster_bidirectional_1997}. These \acp{brnn} behave like traditional \ac{rnn} but infer translations from both forward and reverse sequences, addressing this specific requirement.
The \ac{nn} model was designed to generate a categorical probability distribution for each \emph{tr} over the available \emph{hc} set. These distributions were normalised using a \emph{softmax} function, allowing the selection of the \emph{hc} with the highest probability for each \emph{tr}. In the context of \ac{sh}, an \ac{rnn} model $M$ can be represented as a function: $M: \{tr\} \rightarrow \{hc\}$.

This work aims to leverage the same motivations and strategies from the previous approach but introduce a novel support for \ac{cnn} models for syntax highlighting. The motivation behind this shift lies in the previous approach's near-perfect \ac{sh} accuracy and the inherent parallelisation design of \ac{cnn} models, which can significantly reduce computation time. This transition to \acp{cnn} aims to retain the accuracy achieved by \acp{rnn} while enabling faster computations, particularly essential when dealing with large datasets and multiple programming languages.
\acp{cnn} are well-suited for tasks involving structured data, such as syntax highlighting, due to their ability to capture local patterns efficiently. While more recent techniques have evolved, \acp{cnn} offer a lightweight yet robust solution for this specific application, ensuring stable performance across various programming languages and coverage settings.

\subsection{CNN Model}

\acf{cnn} models have demonstrated their effectiveness in sequence-to-sequence learning tasks, surpassing the capabilities of traditional recurrent models \cite{gehring2017convolutional,nguyen_2016}. One of the key advantages of \ac{cnn}s lies in their inherent ability to enable fully parallelised training, optimising the utilisation of \emph{GPU} hardware. This parallel processing not only enhances training speed but also boosts performance in both training and inference stages \cite{gehring2017convolutional, nguyen_2016}. Additionally, \ac{cnn}s serve as adept feature extractors, capable of discerning meaningful representations even from limited training data. This feature extraction ability not only offers regularisation benefits in tasks with small datasets but also ensures an expanded receptive field on the input. This receptive field grows with the number of layers, enabling the model to effectively capture dependencies across input segments. The benefits of \ac{cnn}s are leveraged to improve the previously established approach and mirror the functionalities of the \ac{rnn}s.

A streamlined \ac{cnn} model, inspired by the model proposed by Ngoc et al. \cite{nguyen_2016}, is introduced for its close alignment with the requirements of the \ac{sh} task. The method operates on the tokenised sequence of the syntax structure, denoted as $X_{in}$, where $X_{in}$ represents the number of input channels. Notably, each element $x \in X$ falls within the range of [0, 256], with each value representing a keyword requiring highlighting. Given that the input data comprises one-dimensional sequences, two one-dimensional convolutional layers are employed to simultaneously capture local and global features. These layers process the textual data to ensure the extraction of essential information. The resulting feature maps from these layers are concatenated and subsequently passed through a series of convolutional layers. To prevent overfitting, dropout regularisation is applied, enhancing the model's generalisation capabilities. Finally, the extracted features are fed into a fully connected layer, converting them into the respective output classes—namely, the \emph{hc}.

\section{Experiments}
\label{sec:experiments}


This paper addresses seven critical research questions that collectively provide a comprehensive understanding of the performance, efficiency, and generalisation capabilities of the \emph{state-of-the-art} \emph{sh} models, with a particular emphasis on the impact of the proposed \ac{cnn} model and \emph{GPU} accelerations.
The first three research questions examine the generalisation of the \emph{state-of-the-art} \ac{rnn} and \ac{brnn} models when applied to an extended dataset, considering the accuracy, prediction delays, and accuracy in handling code snippets.
Following this, the same set of questions is evaluated in the context of the proposed \ac{cnn} models, thus verifying that the proposed models too can provide near-perfect accuracy like \emph{state-of-the-art} models, but with smaller prediction delays.
Finally, the last research question focuses on evaluating the speed-ups that both the \emph{state-of-the-art} \ac{rnn} and the proposed CNN models can experience when evaluated on \emph{GPU}s.

\begin{reqs}
    \item [\req{1}] To what extent does the original \ac{nn}-based approach maintain its near-perfect accuracy when applied to a broader set of mainstream programming languages and various levels of grammatical coverage?
\end{reqs}
This questions evaluates to what extent the \emph{state-of-the-art} approach for \emph{On-the-Fly} \ac{sh} can maintain its near-perfect accuracy score, for any level of coverage, to a new set of programming languages.
\smallskip

\begin{reqs}
    \item [\req{2}] How does the prediction speed of the three \ac{sh} approaches continue to compare on the mainstream programming languages?
\end{reqs}
With the speed of evaluation being an important factor of \emph{On-the-Fly} \ac{sh}, this question provides an overview of the time delays requested by each \ac{sh} resolver.
\smallskip

\begin{reqs}
    \item [\req{3}] Compared to the \ac{re} and \ac{bf} approaches, to what extent can the original \ac{rnn} based approach continue to produce near-perfect \ac{sh} solutions for incorrect or incomplete language derivations, on a new set of programming languages?
\end{reqs}
With online \ac{sh} requiring the highlighting of incorrect languages derivations (such as snippets, diffs, or different language versions), this questions investigates how the originally proposed \ac{nn}-based approach can continue to deliver its accuracy gains on a new set of programming languages.
\smallskip

\begin{reqs}
    \item [\req{4}] How does the proposed \ac{cnn} model compare to the original \ac{rnn} resolvers in terms of \ac{sh} accuracy?
\end{reqs}
With the original approach delivering near-perfect \ac{sh} accuracy, this question aims at evaluating whether the proposed \ac{cnn} model can stack up to these similar figures.
\smallskip

\begin{reqs}
    \item [\req{5}] How does the prediction delays of the proposed \ac{cnn} model compare to the original \ac{nn} resolvers?
\end{reqs}
This question evaluates whether the proposed \ac{cnn} model can indeed provide shorter evaluation delays over the baseline \ac{nn} solution.
\smallskip

\begin{reqs}
    \item [\req{6}] How accurately can the proposed \ac{cnn} models provide \ac{sh} for incorrect or incomplete language derivations?
\end{reqs}
This question reports on the suitability of the proposed approach in providing \ac{sh} for incorrect language derivations, following similar motivations of \req{4}.
\smallskip

\begin{reqs}
    \item [\req{7}] How does the utilisation of \emph{GPU}s impact the prediction delays of the \emph{state-of-the-art} \ac{rnn} and \ac{brnn}, and the proposed \ac{cnn} models?
\end{reqs}
Given the time-sensitive nature of \ac{sh} in online environments, this research question investigates the potential for speed-up gains by harnessing the computational power of \emph{GPU}s. Such question is motivated by the renowned ability of \emph{GPU}s to accelerate the execution of deep learning models such as the \emph{state-of-the-art} \ac{rnn} and \ac{brnn}, and the proposed \ac{cnn} models.
\smallskip

\subsection{Data, Entities and Metrics}
This section outlines the entities and metrics relevant to the experiments, including the models under investigation, training procedures, cross-validation setup, coverage tasks, accuracy evaluation, benchmarks, and snippet evaluation.

\paragraph{Coverage Tasks} In the experiments, \ac{sh} models are validated based on their ability to highlight characters in language derivation, as per three coverage tasks (\emph{T1}, \emph{T2}, \emph{T3}, \emph{T4}). These tasks follow the validation strategy employed in the \emph{state-of-the-art} approach. Each Coverage Task (\emph{T}) is constructed by grouping various language feature categories, each representing a unique \emph{hc}. These categories include lexically identifiable token classes, special types of identifiers, and classes for the classification of token identifiers~\cite{10.1145/3540250.3549109}.
These coverage tasks are designed to evaluate the models' adaptability to different \ac{sh} coverage requirements, ensuring that they can identify and highlight specific language features. The specific criteria for each coverage task involve detecting different feature groups, including identifiers, literals, and annotations, in the code. More details about the language feature groups and their detection strategy can be found in the replication package.

\paragraph{Extended Dataset} The dataset extension process, involving the inclusion of \emph{JavaScript}, \emph{C++}, and \emph{C\#}, closely mirrors the methodology employed in {Palma et. al}\cite{10.1145/3540250.3549109}, with the primary difference being the development of new highlighting \ac{ast} walkers for these three additional programming languages. Just as in the prior approach, the process starts with data mining. GitHub's public repositories are accessed through the GitHub API, and files matching the respective language extensions are collected. The collected files must be those for which the \ac{bf} strategy can derive an \ac{ast}. For each programming language, the data collection continues until a sample size of $20,000$ unique files is reached, with uniqueness here being determined at a token-id sequence level~\cite{10.1145/3540250.3549109}. This sample size allows for comprehensive accuracy and performance testing. The collected datasets for each language include statistics on various aspects, such as the number of characters, whitespace, lines of code, and tokens, as reported in Table \ref{tab:experiments:file_stats}.

\begin{table*}[tb]
    \caption{Metrics for \emph{Java}, \emph{Kotlin}, \emph{Python}, \emph{C++}, \emph{C\#}, and \emph{JavaScript} normalised \emph{SH} oracles}
    \label{tab:experiments:file_stats}
    \centering
    \resizebox{1.0\linewidth}{!}{
        \rowcolors{2}{gray!10}{}
\begin{tabular}{
    l
    S[table-format=5] S[table-format=6] S[table-format=2] S[table-format=4] S[table-format=8]
    S[table-format=5] S[table-format=6] S[table-format=2] S[table-format=4] S[table-format=8]
    S[table-format=5] S[table-format=6] S[table-format=2] S[table-format=4] S[table-format=8]
}

\hiderowcolors
\toprule

\multirow{2}[2]{*}{\textbf{Metric}} & \multicolumn{5}{c}{\textbf{\java}} & \multicolumn{5}{c}{\textbf{\kotlin}} & \multicolumn{5}{c}{\textbf{\python}} \\
\cmidrule(lr){2-6} \cmidrule(lr){7-11} \cmidrule(lr){12-16}
& {\textbf{Mean}} & {\textbf{SD}} & {\textbf{Min}} & {\textbf{Median}} & {\textbf{Max}} & {\textbf{Mean}} & {\textbf{SD}} & {\textbf{Min}} & {\textbf{Median}} & {\textbf{Max}} & {\textbf{Mean}} & {\textbf{SD}} & {\textbf{Min}} & {\textbf{Median}} & {\textbf{Max}} \\

\midrule
\showrowcolors

Chars & 6239 & 11575 & 0 & 2932 & 504059 & 2455 & 4385 & 80 & 1490 & 176176 & 7391 & 34325 & 0 & 3398 & 3987090 \\
Whitespaces & 1207 & 2417 & 0 & 529 & 72702 & 575 & 1276 & 6 & 282 & 47495 & 1999 & 12941 & 0 & 829 & 1465856 \\
Lines & 190 & 332 & 1 & 94 & 14628 & 70 & 121 & 5 & 43 & 4734 & 208 & 873 & 1 & 104 & 89373 \\
Tokens & 882 & 1745 & 1 & 371 & 45229 & 737 & 1559 & 23 & 327 & 72484 & 1161 & 4997 & 1 & 525 & 448562 \setcounter{rownum}{0} \\

\hiderowcolors
\toprule

\textbf{Metric} & \multicolumn{5}{c}{\textbf{\cpp}} & \multicolumn{5}{c}{\textbf{\csharp}} & \multicolumn{5}{c}{\textbf{\javascript}} \\

\midrule
\showrowcolors

Chars & 27095 & 392923 & 0 & 2710 & 23944620 & 7016 & 28002 & 0 & 2325 & 2811507 & 14774 & 132637 & 0 & 1843 & 8937963 \\
Whitespaces & 8793 & 140098 & 0 & 436 & 7160063 & 2199 & 7282 & 0 & 578 & 407133 & 3437 & 34329 & 0 & 428 & 2933810 \\
Lines & 427 & 3762 & 1 & 94 & 222894 & 179 & 442 & 1 & 68 & 13293 & 350 & 3125 & 1 & 62 & 277838 \\
Tokens & 750 & 4082 & 1 & 186 & 222265 & 1253 & 4894 & 1 & 405 & 326685 & 3576 & 33011 & 1 & 405 & 2191766 \\

\bottomrule

\end{tabular}

    }
\end{table*}




The extended dataset adheres to the structure of the original dataset also with regards to the coverage tasks. In fact, each of the language datasets is duplicated four times, with each duplication configuring the \ac{sh} targets to correspond to the specific colours associated with one of the four coverage tasks. This adaptation of highlighting targets is accomplished using the \emph{Task Adapter} concept for which targets for \emph{T1}, \emph{T2}, and \emph{T3} are reductions of the targets or \emph{T4}~\cite{10.1145/3540250.3549109}.

\paragraph{Cross-Validation Setup} All accuracy experiments are validated using a three-fold cross-validation setup. The dataset for each language's coverage task is partitioned into three distinct folds, with each fold consisting of a training, testing and validation dataset.
These splits entail a 33\%-66\% division into testing and training sets, with a 10\% subset of the training data reserved for validation.
Additionally, following the steps of \cite{10.1145/3540250.3549109}, for each fold, the test subset is used as source set for the generation of the 5000 incorrect derivations validation dataset.

\paragraph{Incorrect Derivations} To evaluate the accuracy of \ac{sh} for incomplete (invalid) language derivations, the approach focuses on generating invalid language derivations from a set of valid sampled files. This shift in focus is due to the infeasibility of generating correct SH for incorrect files using the BF method. For this purpose, test files are sampled line-wise to create files of snippet size. The snippet lengths are determined based on the language's mean, standard deviation, minimum, and maximum snippet line numbers, obtained from StackExchange's Data Explorer. For each language, at time of testing these amounted to (mean, standard deviation, minimum, maximum): \num{23.23}, \num{15.00}, \num{1}, and \num{1157} for \emph{JavaScript}; \num{17.00}, \num{28.71}, \num{1}, and \num{1234} for \emph{C++}; and \num{17.00}, \num{26.89}, \num{1}, and \num{1218} for \emph{C\#}. Snippets are drawn randomly from the test files, and the process generates HETAs for these snippets.

\paragraph{Models Under Investigation} The experiment setup aim at investigating the \ac{sh} performances of this set of \ac{sh} models:
\begin{enumerate}
    \item \textbf{Brute Force:} \ac{bf} models, based on \emph{ANTLR4}~\cite{antlr} carried over from prior work for \emph{Java}, \emph{Python}, and \emph{Kotlin}, and introduced anew in this paper for \emph{JavaScript}, \emph{C++} and \emph{C\#}.
    \item \textbf{State-of-Practice:} referring to previous work~\cite{10.1145/3540250.3549109} in this space which considers \emph{Pygments}~\cite{pygments}, all the experiment setups report updated metrics for the latest version at the time of writing \emph{v2.13}
    \item \textbf{State-of-the-Art Models:} These include \ac{rnn} models with different hidden unit sizes (\ac{rnn}16, \ac{rnn}32, \ac{rnn}64) and the all the bidirectional variants (\ac{brnn}16, \ac{brnn}32, \ac{brnn}64).
    \item \textbf{Proposed Models:} These include the \ac{cnn} models introduced by this work (\ac{cnn}32, \ac{cnn}64 and \ac{cnn}128).
\end{enumerate}

\paragraph{Training Procedures} All deep learning models presented in this work (\ac{rnn}, \ac{brnn} and \ac{cnn}) are trained using the same training configuration presented in \emph{Palma et. al}~\cite{10.1145/3540250.3549109}, which instructs about the choice of optimiser, learning rate, batch size, and epoch count. Hence each model is trained sequentially on each training sample, with cross-entropy loss and Adam optimiser. The training session for any SH RNN, language and coverage, was accordingly set to train for two epochs with a learning rate of $10^{-3}$, and for a subsequent two epochs with a learning rate of $10^{-3}$.

\paragraph{Accuracy Metric} Adhering to the methodology of the prior approach, this investigation focuses on ascertaining the highlighters' capacity to associate each character in the input text with the correct \ac{sh} class, for each coverage specification. This approach also serves to reconcile potential discrepancies in tokenisation between the \ac{bf} and Regex strategies for a given input file.

\paragraph{Benchmarks} The prediction speed measurement during the experiments involves evaluating the time delays for each \ac{sh} resolver. This metric quantifies the absolute time in nanoseconds required to predict the \ac{sh} of an input file. The benchmarking encompasses various \ac{sh} methods, and adheres with the measurement techniques of \emph{state-of-the-art}~\cite{10.1145/3540250.3549109}. For each model the following time delays are evaluated:
\begin{itemize}
    \item \textbf{Brute-Force:} Time taken to natively lex and parse the input file and execute a \ac{sh} walk on the acquired \ac{ast}.
    \item \textbf{State-of-Practice:} Time measured for computing the output vector of \ac{sh} classes, given the source text of the file. It excludes the time consumed for formatting the output according to any specific specification to emphasize the intrinsic time complexity of the underlying \ac{sh} strategy.
    \item \textbf{Neural Networks:} This measurement comprises two components. First, the time required for the lexer inherited from \emph{ANTLR4} (the same lexer employed by the BF approach) to tokenize the input file, resulting in a sequence of token rules. Subsequently, it factors in the time for the \ac{nn} model to create the input tensor and predict the complete output vector of \ac{sh} classes.
\end{itemize}



\subsection{Evaluating the Generalisation of the (B)RNN Approach}
\label{sec:experiments:generalisation}

This section addresses the experimental setups for investigation into the generalisation of the \emph{state-of-the-art} \ac{rnn} approach for \ac{sh}. This interests \req{1}, \req{2} and \req{3}. The primary objective is to assess the models' capacity to deliver similar levels of performance in a broader range of programming languages. The performance metrics under scrutiny encompass accuracy, coverage, and the execution time of the models.

To accomplish this, an extended dataset was curated to include three additional mainstream programming languages: \emph{JavaScript}, \emph{C++}, and \emph{C\#}. This expansion followed the strategy introduced in the previous approach, focusing on the creation of language-specific oracle models based on the languages' original lexers and parsers, relying on the \emph{ANTLR4}~\cite{antlr} grammars for each language.


Following the original training configuration, \ac{rnn}16, \ac{rnn}32, \ac{brnn}16 and \ac{brnn}32 models were trained for each coverage task for every new language. In particular, these were trained on the training set of each task fold.
Furthermore, trainings were performed on the same hardware configuration originally used.

In the evaluation phase, the performance of the resulting \ac{rnn} models, the \emph{state-of-practice}, and \ac{bf} resolvers were examined across several dimensions. For assessing accuracy, each \ac{rnn} model was tested on the dedicated test dataset corresponding to the fold on which it was trained. Similarly, the \emph{state-of-practice} resolvers underwent accuracy testing on the same fold test datasets. Notably, the \ac{bf} method, which is known for producing perfect predictions, was not included in accuracy testing. For the evaluation of incorrect derivations, all resolvers, including \ac{bf}, were subjected to testing on the incomplete test dataset for each fold. In this context, the resolution process remained consistent across all methods. Speed benchmarks were recorded differently, with timing measurements conducted for each resolver on the entire dataset consisting of $20,000$ files for each programming language and averaged across 30 reruns per file.
Such conditions, together with the details discussed earlier, ensure that the models are trained and evaluated under the same circumstances as those applied in the evaluation of the \emph{state-of-the-art} approach~\cite{10.1145/3540250.3549109}.

\subsection{Evaluating the CNN Models}

This set of experiments aims to assess the effectiveness of the newly proposed \ac{cnn} resolvers in achieving \ac{sh} accuracy on par with or exceeding the \ac{rnn} solutions, while achieving faster prediction times. The experiments are designed to address the research questions \req{4}, \req{5}, and \req{6}.

The evaluation follows the same validation processed carried out by \emph{state-of-the-art} on the \ac{rnn} approach, utilising the extended dataset previously described in Section \ref{sec:experiments:generalisation}, which covers languages such as \emph{Java}, \emph{JavaScript}, \emph{Kotlin}, \emph{Python}, \emph{C++}, and \emph{C\#}. The proposed \ac{cnn} models, specifically \emph{CNN64} and \emph{CNN128}, are trained using the same training configuration originally developed for the \ac{rnn} and \ac{brnn} models.

In the preliminary analysis, hyperparameters were investigated using the \emph{Java} validation dataset as a reference, with the objective of determining optimal settings for the \ac{cnn} model. This exploration revealed that smaller kernel sizes, specifically values of 3, 5, and 7, produced the most effective results for the defined objectives. The choice of smaller kernel sizes was informed by the immediate relationships between elements in the sequence, which are characteristic of the short-term connections prevalent in the \ac{sh} task. Unlike translation tasks that heavily rely on long-term dependencies between words, the nature of syntax highlighting called for a more nuanced approach, favouring smaller kernel sizes. The exploration also involved varying the number of layers in the \ac{cnn} stack, from 1 to 4 layers, with fewer layers emerging as the superior choice, likely due to the unique demands of the task. Consistency was maintained by using the same hidden and embedding dimensions employed in the previous RNN models, with dimensions ranging from $2^4$ to $2^8$. To simplify the exploration, different dropout values were tested within a range of 0.1 to 0.5. Initially, the embedding size was fixed, and then systematically increased. Subsequent investigations involved expanding hidden dimensions and other parameters. Three model configurations, where performance converged, were selected for thorough evaluation. Hence this set of experiments evaluates the following models: \ac{cnn}32 features a single layer with 32 hidden units and a 32-layer embedding, \ac{cnn}64 features one layer with 64 hidden units and a 64-layer embedding, and \ac{cnn}128 features a single layer with 128 hidden units and a 128-layer embedding.

The evaluation of the newly introduced \ac{cnn} models follows a comprehensive approach. These \ac{cnn} models were trained on all six programming languages within the extended dataset, and for each of the four defined coverage tasks, similar to the \ac{rnn} models. Accuracy testing was conducted on the complete and incorrect files, mirroring the evaluation process of the \ac{rnn} models. Additionally, the benchmarking of \ac{cnn} models was carried out in the same manner as the \ac{rnn} models. It is important to note that the \ac{cnn} models underwent training and benchmarking on the same hardware configuration used for the \ac{rnn} models. This consistency in testing procedures and hardware configuration facilitates a direct and meaningful comparison between the \emph{state-of-the-art} \ac{rnn} models and the novel \ac{cnn} models.


\subsection{Evaluating GPU Speed-ups}

This experimental setup interest \req{7}, and focus on the evaluation of execution speed improvements attained by employing \emph{GPU}s for \ac{sh} prediction. It includes a comparison between the \emph{state-of-the-art} \ac{rnn} models (\ac{rnn}16, \ac{rnn}32, \ac{brnn}16 and \ac{brnn}32) and the proposed \ac{cnn} models (\ac{cnn}32, \ac{cnn}64, \ac{cnn}128) concerning prediction speed when using a \emph{GPU}.

For each of the six languages in the extended dataset (\emph{Java}, \emph{JavaScript}, \emph{Kotlin}, \emph{Python}, \emph{C++}, and \emph{C\#}), \ac{sh} predictions were conducted on each of the $20,000$ files contained in the respective language dataset. As the speed evaluations conducted in~\cite{10.1145/3540250.3549109}, these predictions were repeated 30 times for robustness and consistent evaluation.

The time delays for \ac{sh} prediction take into account both lexing and model prediction, mirroring the parameters used in the previous \emph{state-of-the-art} approach. However, in this set of experiments, the crucial difference lies in the execution of model evaluation on a GPU. Further implementation details and information are provided in the associated replication package.


\subsection{Execution Setup}
All \ac{rnn} and \ac{cnn} models are trained on a machine equipped with an AMD EPYC 7702 \num{64}-Core CPU clocked at \qty{2.00}{\giga\hertz}, \qty{64}{\giga\byte} of RAM, and a single Nvidia Tesla T4 GPU with \qty{16}{\giga\byte} of memory. The same machine is utilised for GPU benchmarking experiments. Instead, all performance testing for all of the compared approaches was carried out on the same machine with an 8-Core Intel Xeon(R) Gold 6126 CPU clocked at 2.60GHz with 62 GB of RAM.

\subsection{Threats to Validity}
\label{subsec:experiments:threats}

The adoption of \textsc{ANTLR4} as a unified framework for defining and evaluating the \ac{bf} model in the context of real-time \ac{sh} presents a well-rounded choice. However, the existence of alternative parsing tools, some of which might be tailored to specific programming languages, could potentially influence the efficacy of \ac{bf} resolvers. The selection of such tools should be aligned with the practical demands of online \ac{sh}, as outlined in \cref{sec:introduction}.

A notable challenge is the reliance on synthetically generated, incomplete or incorrect language constructs to meet requirements \req{3} and \req{6}. This synthetic approach, while practical, lacks the direct correlation with real-world user-generated code snippets, necessitating a cautious interpretation of results. Despite this, the synthetic dataset serves to indirectly verify the model's capacity to deduce likely missing contexts, although it introduces a degree of variability inherent to its manual creation process, thus underlining the reliance on statistical approximations.

This study's comparison with \pygments, which supports a vast array more than \num{500} languages, adds significant value. Nevertheless, the limitation stems from comparing only a subset of languages (\java, \kotlin, \python, \emph{C++}, \emph{C\#}, and \emph{JavaScript}), suggesting broader applicability through language-specific \ac{bf} training. A more comprehensive evaluation across all languages supported by \apprregex-based alternatives would enhance the understanding of the proposed approach's abilities.

Moreover, the predictive delay benchmarks, while providing an overview of tool performance, might not fully capture nuances related to specific implementation choices or factors inherent to different platforms, such as online file size constraints. Aspects like integration effectiveness, caching mechanisms, and hardware capabilities could also influence the performance of \ac{sh} solutions. The efficiency of the suggested \ac{cnn} approach might vary across different operational environments, such as when deployed using advanced deep learning frameworks~\cite{abadi_tensorflow_2015} or on GPU hardware, suggesting potential avenues for further investigation and optimisation.

\section{Results}
Expanding upon the experimental configurations detailed in Section \ref{sec:experiments}, this section delves into a comprehensive analysis of the proposed approach's performance in response to the four research questions outlined. To facilitate comparisons, the \enquote{Kruskal-Wallis H} test \cite{montgomery_design_2017} was employed in tandem with the \enquote{Vargha-Delaney $\hat{A}_{12}$} test~\cite{vargha_critique_2000} to gauge the effect size, shedding light on the magnitude of observed differences. Consequently, the ensuing discussion presents the evaluation metrics in terms of median values, a choice motivated by the tests' foundation in assessing median differences.

\subsection{\req{1} -- Generalisation: Accuracy}
\label{subsec:results:rq1}
In response to \req{1}, which examines the ability of the original \ac{nn} based approach to retain its near-perfect accuracy when applied to a broader set of mainstream programming languages and various levels of grammatical coverage, significant insights were obtained. Table \ref{tab:results:accuracy} summarises the results for the \ac{sh} accuracy obtained by each resolver, in all combinations of task and language.


The examination of generalisation performance for \ac{sh} accuracy in \emph{C++} reveals intriguing findings. The base \ac{rnn} models, \ac{rnn}16 and \ac{rnn}32, continue to exhibit accuracy gains akin to those observed in previous research involving \emph{Java}, \emph{Kotlin}, and \emph{Python}. However, these models display slightly higher accuracy in \emph{C++}. Their accuracy scores approach near-perfection, with exceptions noted in scenarios where correct highlighting hinges on deterministically feasible token look-ahead.

The bidirectional variants, namely \ac{brnn}16 and \ac{brnn}32, consistently deliver near-perfect accuracy scores. Surprisingly, the \emph{state-of-the-art} resolvers demonstrate a slightly improved performance in \emph{C++} compared to \emph{Java}, \emph{Kotlin}, and \emph{Python}. However, they still remain significantly inferior to the \ac{nn} based solutions. Furthermore, the distribution of accuracy scores is notably more compact and skewed towards perfect accuracy in both the \ac{rnn} and \ac{brnn} models, in contrast to the \emph{state-of-the-art} solution, which exhibits larger variance in its predictions. This pattern aligns with previous observations in \emph{Java}, \emph{Kotlin}, and \emph{Python}, affirming the approach's capacity to generalize in terms of accuracy to \emph{C++}.


The results for \emph{C\#} mirror the conclusions drawn from the \emph{C++} analysis. The \ac{rnn} models, while not performing as remarkably as in \emph{C++}, demonstrate accuracy levels closer to those observed in \emph{Java}, \emph{Kotlin}, and \emph{Python}. Therefore, the approach maintains its ability to generalize its accuracy to \emph{C\#}.


JavaScript showcases a unique scenario. Accuracy scores for the \ac{rnn} models in this language are the lowest among all six languages, with median values hovering in the low 90s. However, the accuracy of the state-of-practice resolvers is consistent with what has been observed in other languages. The bidirectional networks, on the other hand, continue to deliver near-perfect accuracy. Similar to other languages, the distribution of accuracy scores is more densely concentrated towards perfect accuracy in both the \ac{rnn}s and even more so in the \ac{brnn}s, compared to the \emph{state-of-the-art} approach. It is worth noting that base \ac{rnn} models exhibit a small number of results below 50\% accuracy, an anomaly not observed in the other five languages.

Overall, the accuracy of the state-of-practice resolvers remains consistent with previous research conducted on \emph{Java}, \emph{Kotlin}, and \emph{Python}. These resolvers, however, consistently perform worse than all \ac{rnn} and \ac{brnn} models. While a slight drop in accuracy was observed for \emph{JavaScript}, this issue is not present in the non-baseline bidirectional networks, which continue to deliver near-perfect performances.

\begin{table*}[tb]
    \caption{Median values over \num{3} folds for the accuracy. The maximum scores per task are highlighted}
    \label{tab:results:accuracy}
    \centering
    \sisetup{table-format=1.4}
\rowcolors{2}{gray!10}{}
\begin{tabular}{
    l SSSS SSSS SSSS
}

\hiderowcolors
\toprule

\multirow{2}[2]{*}{\textbf{Model}} & \multicolumn{4}{c}{\textbf{\java}} & \multicolumn{4}{c}{\textbf{\kotlin}} & \multicolumn{4}{c}{\textbf{\python}} \\
\cmidrule(lr){2-5} \cmidrule(lr){6-9} \cmidrule(lr){10-13}
& {\textbf{\task{1}}} & {\textbf{\task{2}}} & {\textbf{\task{3}}} & {\textbf{\task{4}}} & {\textbf{\task{1}}} & {\textbf{\task{2}}} & {\textbf{\task{3}}} & {\textbf{\task{4}}} & {\textbf{\task{1}}} & {\textbf{\task{2}}} & {\textbf{\task{3}}} & {\textbf{\task{4}}} \\

\midrule
\showrowcolors

\apprregex{} & 0.8649 & 0.7610 & 0.7243 & 0.7240 & 0.7949 & 0.6944 & 0.6733 & 0.6718 & 0.9339 & 0.8163 & 0.8163 & 0.8141 \\
\apprrnn{16} & 0.9987 & 0.9716 & 0.9676 & 0.9668 & \tabhvalue 1.0000 & 0.9627 & 0.9598 & 0.9604 & \tabhvalue 1.0000 & 0.9560 & 0.9559 & 0.9550 \\
\apprrnn{32} & \tabhvalue 1.0000 & 0.9751 & 0.9710 & 0.9706 & \tabhvalue 1.0000 & 0.9648 & 0.9640 & 0.9630 & \tabhvalue 1.0000 & 0.9572 & 0.9571 & 0.9570 \\
\apprbrnn{16} & \tabhvalue 1.0000 & \tabhvalue 1.0000 & \tabhvalue 1.0000 & \tabhvalue 1.0000 & \tabhvalue 1.0000 & \tabhvalue 1.0000 & \tabhvalue 1.0000 & \tabhvalue 1.0000 & \tabhvalue 1.0000 & \tabhvalue 1.0000 & \tabhvalue 1.0000 & \tabhvalue 1.0000 \\
\apprbrnn{32} & \tabhvalue 1.0000 & \tabhvalue 1.0000 & \tabhvalue 1.0000 & \tabhvalue 1.0000 & \tabhvalue 1.0000 & \tabhvalue 1.0000 & \tabhvalue 1.0000 & \tabhvalue 1.0000 & \tabhvalue 1.0000 & \tabhvalue 1.0000 & \tabhvalue 1.0000 & \tabhvalue 1.0000 \\
\apprcnn{32} & \tabhvalue 1.0000 & \tabhvalue 1.0000 & \tabhvalue 1.0000 & \tabhvalue 1.0000 & \tabhvalue 1.0000 & \tabhvalue 1.0000 & \tabhvalue 1.0000 & \tabhvalue 1.0000 & \tabhvalue 1.0000 & \tabhvalue 1.0000 & \tabhvalue 1.0000 & \tabhvalue 1.0000 \\
\apprcnn{64} & \tabhvalue 1.0000 & \tabhvalue 1.0000 & \tabhvalue 1.0000 & \tabhvalue 1.0000 & \tabhvalue 1.0000 & \tabhvalue 1.0000 & \tabhvalue 1.0000 & \tabhvalue 1.0000 & \tabhvalue 1.0000 & \tabhvalue 1.0000 & \tabhvalue 1.0000 & \tabhvalue 1.0000 \\
\apprcnn{128} & \tabhvalue 1.0000 & \tabhvalue 1.0000 & \tabhvalue 1.0000 & \tabhvalue 1.0000 & \tabhvalue 1.0000 & \tabhvalue 1.0000 & \tabhvalue 1.0000 & \tabhvalue 1.0000 & \tabhvalue 1.0000 & \tabhvalue 1.0000 & \tabhvalue 1.0000 & \tabhvalue 1.0000 \setcounter{rownum}{0} \\

\hiderowcolors
\toprule

\textbf{Model} & \multicolumn{4}{c}{\textbf{\cpp}} & \multicolumn{4}{c}{\textbf{\csharp}} & \multicolumn{4}{c}{\textbf{\javascript}} \\

\midrule
\showrowcolors

\apprregex{} & 0.8977 & 0.9513 & 0.8703 & 0.8703 & 0.8840 & 0.7520 & 0.7284 & 0.7284 & 0.9450 & 0.8266 & 0.7991 & 0.7991 \\
\apprrnn{16} & 0.9972 & \tabhvalue 1.0000 & 0.9893 & 0.9915 & 0.9901 & 0.9580 & 0.9468 & 0.9463 & \tabhvalue 1.0000 & 0.9236 & 0.9186 & 0.9242 \\
\apprrnn{32} & 0.9982 & \tabhvalue 1.0000 & 0.9953 & 0.9953 & 0.9913 & 0.9646 & 0.9608 & 0.9603 & \tabhvalue 1.0000 & 0.9000 & 0.9309 & 0.9299 \\
\apprbrnn{16} & \tabhvalue 1.0000 & \tabhvalue 1.0000 & \tabhvalue 1.0000 & \tabhvalue 1.0000 & \tabhvalue 1.0000 & \tabhvalue 1.0000 & \tabhvalue 1.0000 & \tabhvalue 1.0000 & \tabhvalue 1.0000 & \tabhvalue 1.0000 & \tabhvalue 1.0000 & \tabhvalue 1.0000 \\
\apprbrnn{32} & \tabhvalue 1.0000 & \tabhvalue 1.0000 & \tabhvalue 1.0000 & \tabhvalue 1.0000 & \tabhvalue 1.0000 & \tabhvalue 1.0000 & \tabhvalue 1.0000 & \tabhvalue 1.0000 & \tabhvalue 1.0000 & \tabhvalue 1.0000 & \tabhvalue 1.0000 & \tabhvalue 1.0000 \\
\apprcnn{32} & \tabhvalue 1.0000 & \tabhvalue 1.0000 & \tabhvalue 1.0000 & \tabhvalue 1.0000 & \tabhvalue 1.0000 & 0.9986 & 0.9979 & 0.9979 & \tabhvalue 1.0000 & \tabhvalue 1.0000 & \tabhvalue 1.0000 & \tabhvalue 1.0000 \\
\apprcnn{64} & \tabhvalue 1.0000 & \tabhvalue 1.0000 & \tabhvalue 1.0000 & \tabhvalue 1.0000 & \tabhvalue 1.0000 & 0.9989 & 0.9985 & 0.9985 & \tabhvalue 1.0000 & \tabhvalue 1.0000 & \tabhvalue 1.0000 & \tabhvalue 1.0000 \\
\apprcnn{128} & \tabhvalue 1.0000 & \tabhvalue 1.0000 & \tabhvalue 1.0000 & \tabhvalue 1.0000 & \tabhvalue 1.0000 & 0.9992 & 0.9986 & 0.9986 & \tabhvalue 1.0000 & \tabhvalue 1.0000 & \tabhvalue 1.0000 & \tabhvalue 1.0000 \\

\bottomrule

\end{tabular}

\end{table*}

\begin{figure*}[tb]
    \centering
    \includegraphics[width=1.0\linewidth]{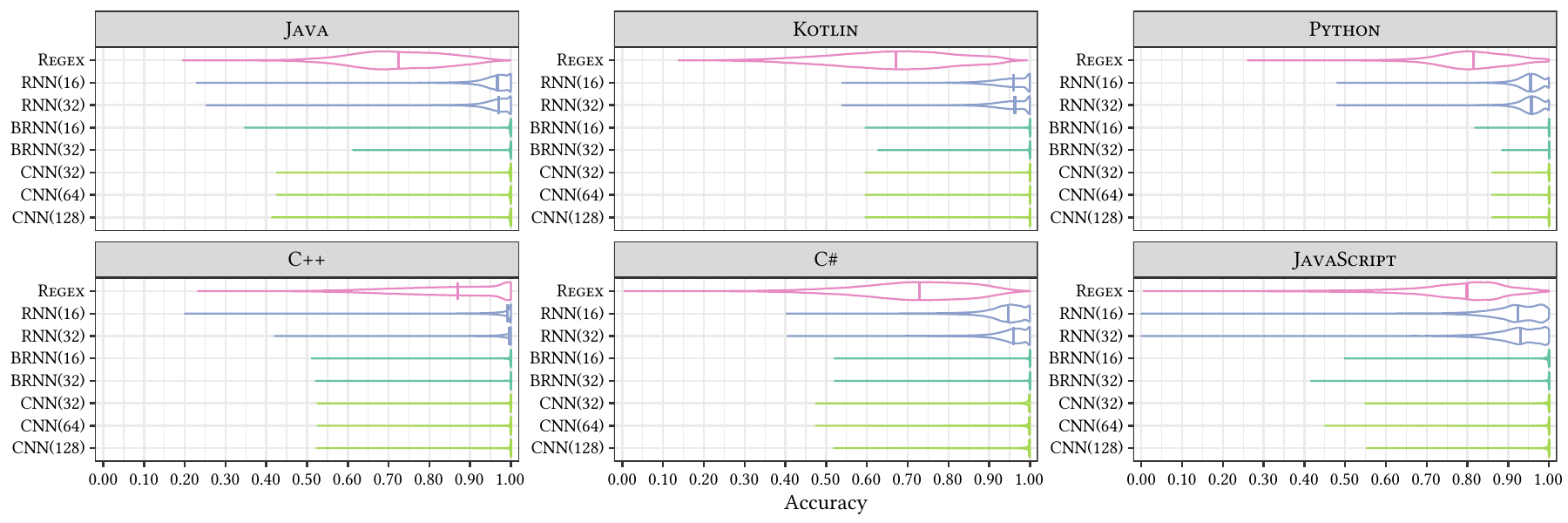}
    \caption{Accuracy values comparison for \task{4}.}
    \label{fig:results:boxplot_accuracy_t4}
\end{figure*}

\begin{table*}[tb]
    \caption{Descriptive statistics of execution time (ms)}
    \label{tab:results:time}
    \centering
    \resizebox{1.0\linewidth}{!}{
    \sisetup{table-format=6.3}
\rowcolors{2}{gray!10}{}
\begin{tabular}{
    l
    S[table-format=3.3] S[table-format=4.3] S[table-format=1.3] S[table-format=2.3] S[table-format=6.3]
    S[table-format=3.3] S[table-format=4.3] S[table-format=1.3] S[table-format=2.3] S[table-format=6.3]
    S[table-format=3.3] S[table-format=4.3] S[table-format=1.3] S[table-format=2.3] S[table-format=6.3]
}

\hiderowcolors
\toprule

\multirow{2}[2]{*}{\textbf{Model}} & \multicolumn{5}{c}{\textbf{\java}} & \multicolumn{5}{c}{\textbf{\kotlin}} & \multicolumn{5}{c}{\textbf{\python}} \\
\cmidrule(lr){2-6} \cmidrule(lr){7-11} \cmidrule(lr){12-16}
& {\textbf{Mean}} & {\textbf{SD}} & {\textbf{Min}} & {\textbf{Median}} & {\textbf{Max}} & {\textbf{Mean}} & {\textbf{SD}} & {\textbf{Min}} & {\textbf{Median}} & {\textbf{Max}} & {\textbf{Mean}} & {\textbf{SD}} & {\textbf{Min}} & {\textbf{Median}} & {\textbf{Max}} \\

\midrule
\showrowcolors

\apprbf{} & 243.748 & 935.827 & 0.004 & 50.703 & 48970.153 & 38.045 & 109.474 & 0.011 & 9.196 & 17129.873 & 55.673 & 252.924 & 0.034 & 25.839 & 27723.622 \\
\apprregex{} & 0.019 & 0.023 & 0.010 & 0.015 & 3.340 & 0.017 & 0.039 & 0.010 & 0.014 & 4.335 & 0.020 & 0.037 & 0.010 & 0.015 & 4.648 \\
\apprrnn{16} & 15.754 & 22.715 & 0.347 & 10.795 & 665.778 & 16.416 & 38.120 & 0.612 & 10.839 & 14940.728 & 81.635 & 297.425 & 0.367 & 44.111 & 28530.784 \\
\apprrnn{32} & 17.030 & 24.138 & 0.366 & 11.061 & 806.670 & 17.588 & 36.850 & 0.639 & 11.153 & 13275.436 & 84.829 & 306.712 & 0.363 & 46.744 & 30926.886 \\
\apprgrnn{16} & 11.126 & 21.158 & 0.434 & 4.946 & 570.257 & 11.218 & 35.648 & 0.655 & 5.195 & 14492.819 & 72.075 & 318.203 & 0.420 & 33.962 & 29676.576 \\
\apprgrnn{32} & 11.113 & 21.167 & 0.426 & 4.930 & 585.709 & 10.744 & 35.921 & 0.639 & 4.899 & 14572.781 & 80.569 & 358.720 & 0.398 & 38.057 & 34206.802 \\
\apprbrnn{16} & 26.956 & 42.556 & 0.450 & 16.300 & 1112.187 & 27.417 & 54.053 & 0.917 & 16.360 & 14708.969 & 99.543 & 357.392 & 0.417 & 54.058 & 36217.433 \\
\apprbrnn{32} & 28.833 & 44.477 & 0.462 & 18.013 & 1250.126 & 29.424 & 55.613 & 0.963 & 17.883 & 14902.277 & 102.585 & 360.601 & 0.430 & 55.232 & 36216.305 \\
\apprgbrnn{16} & 21.501 & 41.888 & 0.544 & 9.303 & 1152.935 & 19.356 & 49.915 & 0.905 & 8.573 & 15039.211 & 87.273 & 398.539 & 0.465 & 40.864 & 39490.076 \\
\apprgbrnn{32} & 21.582 & 42.024 & 0.539 & 9.343 & 1145.865 & 18.980 & 48.418 & 0.884 & 8.534 & 14283.623 & 86.608 & 377.965 & 0.477 & 40.706 & 35263.657 \\
\apprcnn{32} & 19.225 & 12.019 & 0.673 & 19.746 & 93.318 & 18.431 & 28.639 & 0.753 & 18.613 & 14284.947 & 77.395 & 268.063 & 0.811 & 47.581 & 28522.633 \\
\apprcnn{64} & 20.191 & 13.195 & 0.701 & 20.160 & 453.234 & 19.225 & 30.516 & 0.838 & 18.542 & 14470.209 & 72.955 & 244.924 & 0.768 & 47.048 & 24690.476 \\
\apprcnn{128} & 21.583 & 14.318 & 0.715 & 20.537 & 223.501 & 20.321 & 30.983 & 0.904 & 18.623 & 14305.525 & 74.613 & 251.771 & 0.883 & 48.048 & 25858.651 \\
\apprgcnn{32} & 0.789 & 0.405 & 0.514 & 0.659 & 40.768 & 1.632 & 28.127 & 0.522 & 0.825 & 14150.149 & 64.155 & 284.315 & 0.539 & 30.500 & 26769.611 \\
\apprgcnn{64} & 0.747 & 0.329 & 0.502 & 0.642 & 39.996 & 1.627 & 28.251 & 0.522 & 0.814 & 14053.773 & 64.444 & 285.610 & 0.525 & 30.616 & 26863.922 \\
\apprgcnn{128} & 0.742 & 0.331 & 0.502 & 0.634 & 40.744 & 1.653 & 29.485 & 0.523 & 0.810 & 14610.805 & 64.237 & 284.855 & 0.497 & 30.535 & 26817.085 \setcounter{rownum}{0} \\

\hiderowcolors
\toprule

\textbf{Model} & \multicolumn{5}{c}{\textbf{\cpp}} & \multicolumn{5}{c}{\textbf{\csharp}} & \multicolumn{5}{c}{\textbf{\javascript}} \\

\midrule
\showrowcolors

\apprbf{} & 34.598 & 119.267 & 0.001 & 7.281 & 5287.692 & 7.766 & 121.491 & 0.011 & 0.904 & 11891.983 & 150.429 & 1084.771 & 0.026 & 16.425 & 91833.385 \\
\apprregex{} & 0.034 & 0.296 & 0.010 & 0.015 & 63.802 & 0.023 & 0.045 & 0.010 & 0.016 & 9.708 & 0.030 & 0.152 & 0.010 & 0.015 & 27.014 \\
\apprrnn{16} & 11.038 & 43.606 & 0.237 & 5.280 & 2589.176 & 19.128 & 52.872 & 0.396 & 11.078 & 3712.954 & 141.235 & 1177.542 & 0.308 & 25.590 & 82025.300 \\
\apprrnn{32} & 12.841 & 47.249 & 0.244 & 5.601 & 2721.487 & 20.814 & 55.376 & 0.393 & 11.623 & 3529.568 & 138.989 & 1135.261 & 0.308 & 26.481 & 79816.063 \\
\apprgrnn{16} & 10.176 & 56.718 & 0.340 & 2.774 & 3915.631 & 15.811 & 60.962 & 0.467 & 5.420 & 4249.555 & 144.089 & 1329.131 & 0.419 & 17.508 & 89162.290 \\
\apprgrnn{32} & 10.136 & 56.839 & 0.346 & 2.768 & 3981.607 & 15.837 & 60.704 & 0.459 & 5.446 & 4196.093 & 143.593 & 1331.705 & 0.396 & 17.336 & 89539.529 \\
\apprbrnn{16} & 19.435 & 85.809 & 0.294 & 8.247 & 5057.532 & 33.646 & 103.206 & 0.499 & 16.840 & 7244.955 & 180.193 & 1496.569 & 0.357 & 32.483 & 105668.888 \\
\apprbrnn{32} & 22.044 & 91.434 & 0.294 & 9.356 & 5194.802 & 36.119 & 108.729 & 0.536 & 18.625 & 7226.380 & 187.685 & 1530.609 & 0.387 & 33.979 & 103981.296 \\
\apprgbrnn{16} & 19.533 & 109.116 & 0.399 & 5.087 & 7759.243 & 30.607 & 119.945 & 0.617 & 10.200 & 8276.269 & 185.262 & 1720.401 & 0.446 & 22.015 & 119257.848 \\
\apprgbrnn{32} & 19.363 & 107.666 & 0.412 & 5.082 & 7598.733 & 30.730 & 120.494 & 0.617 & 10.213 & 9268.634 & 204.786 & 1881.986 & 0.455 & 24.408 & 141257.846 \\
\apprcnn{32} & 20.206 & 14.235 & 0.536 & 19.720 & 625.615 & 19.568 & 13.550 & 0.676 & 19.910 & 670.424 & 109.029 & 870.394 & 0.694 & 30.075 & 61679.980 \\
\apprcnn{64} & 21.607 & 16.201 & 0.568 & 20.531 & 766.439 & 20.584 & 15.803 & 0.718 & 20.181 & 776.559 & 108.286 & 863.380 & 0.768 & 30.672 & 61510.434 \\
\apprcnn{128} & 22.881 & 21.023 & 0.621 & 20.865 & 1158.937 & 22.309 & 21.045 & 0.773 & 20.704 & 1766.114 & 114.549 & 941.239 & 0.752 & 32.341 & 80964.860 \\
\apprgcnn{32} & 1.004 & 4.861 & 0.487 & 0.649 & 301.837 & 1.003 & 1.390 & 0.515 & 0.728 & 140.178 & 108.637 & 999.694 & 0.529 & 13.740 & 67784.865 \\
\apprgcnn{64} & 1.004 & 4.774 & 0.480 & 0.648 & 311.590 & 0.984 & 1.370 & 0.510 & 0.721 & 138.915 & 96.105 & 877.135 & 0.534 & 12.353 & 58771.929 \\
\apprgcnn{128} & 0.986 & 4.720 & 0.451 & 0.633 & 289.385 & 0.996 & 1.391 & 0.519 & 0.728 & 133.209 & 98.953 & 907.218 & 0.500 & 12.678 & 61571.581 \\

\bottomrule

\end{tabular}

    }
\end{table*}

\subsection{\req{2} -- Generalisation: Benchmarking}
\label{subsec:results:rq2}

\req{2} delves into the generalisation of prediction speed for \ac{rnn} and \ac{brnn} models across mainstream programming languages \emph{Java}, \emph{Kotlin}, \emph{Python}, \emph{C++}, \emph{C\#}, and \emph{JavaScript}. The focus is on identifying whether the performance characteristics, particularly the instantaneous response time ~\cite{seow_designing_2008}, observed in prior studies on \emph{Java}, \emph{Kotlin}, and \emph{Python} continue to hold across the expanded set of languages.

To assess this, \ac{rnn} and \ac{brnn} models, including \ac{rnn}16, \ac{rnn}32, \ac{brnn}16, and \ac{brnn}32, were retrained on all six languages. The models trained for \emph{T4} were benchmarked 30 times on each of the 20\emph{k} files in each language's dataset. These experiments were conducted on the same machine, with no \emph{GPU} utilisation for \emph{NN}-based resolver evaluations, ensuring consistency and comparability. The results are summarised in Table \ref{tab:results:time}.

Building on prior work that classified \ac{rnn} and \ac{brnn} prediction delays as within the \emph{instantaneous} response-time category ~\cite{seow_designing_2008}, this study confirms their continued efficiency in this regard. In the instantaneous category, interactions are expected to complete within 100-200 ms ~\cite{dabrowski_40_2011, seow_designing_2008}, aligning with typical user actions like clicking and typing. Additionally, speed-ups of \ac{rnn}s and \ac{brnn}s over BFs in \emph{Java} and \emph{Kotlin}, with comparable efficiency for \emph{Python}, were previously identified and are consistent in this expanded study.

Specifically, \ac{rnn} and \ac{brnn} models consistently fall within the instantaneous category, offering notable speed-ups over the \ac{bf} across languages. \ac{rnn}16 and \ac{rnn}32 demonstrate speed-ups of 15 times for \emph{Java}, 2 times for \emph{Kotlin}, 3 times for \emph{C++}, and at least 1 time faster for \emph{JavaScript}. While performing on par with the \ac{bf} for \emph{Python}, \ac{rnn} models are within the lower bound of the instantaneous category for \emph{C\#}. Similarly, \ac{brnn}16 and \ac{brnn}32 fall within the instantaneous bounds, providing speed-ups over the \ac{bf} of 9 times for \emph{Java}, performing on par for \emph{Kotlin}, and 2 times for \emph{C++}. However, \ac{brnn}s are 2 times slower than the \ac{bf} for \emph{Python}, at least 4 times slower for \emph{C\#}, and on par with \emph{JavaScript}.

\begin{figure*}[tb]
    \centering
    \includegraphics[width=1.0\linewidth]{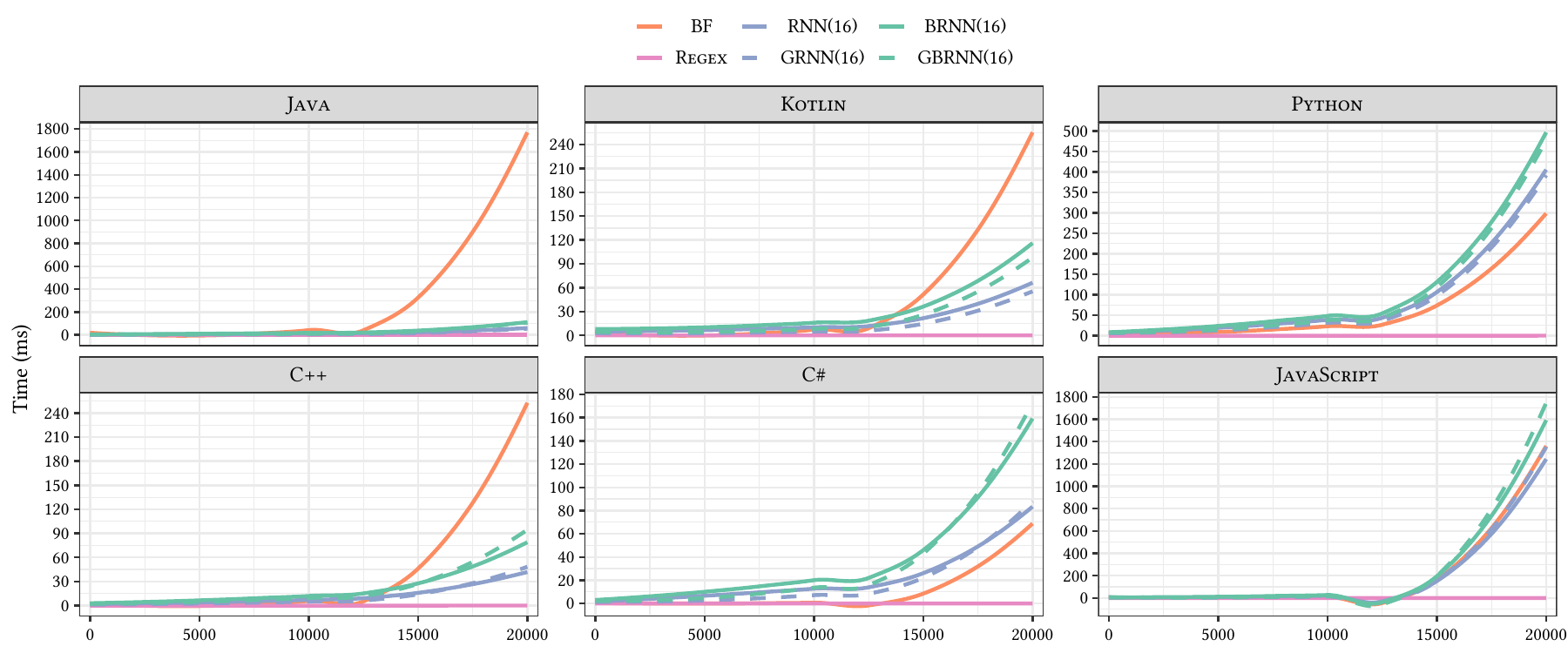}
    \caption{Execution time (ms) values trends comparison for \task{4}.}
    \label{fig:results:lineplot_time}
\end{figure*}

Overall, \ac{rnn} and \ac{brnn} \ac{sh} resolvers consistently operate within the instantaneous response-time category, delivering speed-ups over the \ac{bf} in most cases. Exceptions exist where the \ac{bf} proves to be time-wise efficient, particularly in scenarios with smaller file sizes. This scalability advantage of \ac{rnn} and \ac{brnn} resolvers is evident when compared to the \ac{bf} resolver, as illustrated in Figure \ref{fig:results:lineplot_time}.

\subsection{\req{3} -- Generalisation: Accuracy on Invalid Derivations}
\label{subsec:results:rq3}
Addressing \req{3}, the investigation delved into the \ac{sh} accuracy of \ac{rnn} and \ac{brnn} highlighters confronted with incomplete or incorrect language derivations. Similar to \req{1}, all \ac{rnn} and \ac{brnn} approaches were configured to generate highlighting for all six programming languages and four coverage tasks, with results averaged across three folds. The dataset utilised for \req{3} is the generated snippet dataset, where perfect target solutions are known. The results are summarised in Table \ref{tab:results:accuracy_snippets}.
The findings reveal that the \ac{rnn}-based approaches effectively sustain accuracy performances comparable to those achieved on language derivations where an \ac{ast} is derivable. Significantly, the accuracy values observed for \emph{Java}, \emph{Kotlin}, and \emph{Python} extend to \emph{C++}, \emph{C\#}, and \emph{JavaScript}.
For \ac{rnn}16, across all four tasks, the model exhibits average median accuracies of 96.49\% for \emph{Java}, 96.03\% for\emph{Kotlin}, and 96.95\% for \emph{Python}. Notably, it achieves accuracies of 99.69\% for\emph{C++}, 99.38\% for \emph{C\#}, and 93.95\% for \emph{JavaScript} on the new dataset. Similarly, \emph{\ac{rnn}}32 demonstrates accuracy rates of 96.85\% for \emph{Java}, 93.35\% for\emph{Kotlin}, and 97.11\% for \emph{Python}, while maintaining accuracies of 99.87\% for\emph{C++}, 100\% for \emph{C\#}, and 93.42\% for \emph{JavaScript}.
Both \ac{rnn}16 and \ac{rnn}32 consistently produce near-perfect accuracies for all six languages across all tasks. Furthermore, all models significantly outperform the Regex resolvers across all languages and tasks. While the near-perfect behaviour of the \ac{bf} strategy recorded for \emph{Python} continues for \emph{C++} and \emph{C\#}, deviations for \emph{JavaScript} in \emph{T2}, \emph{T3}, and \emph{T4} are attributed to the snippet strategy employed in this evaluation.
In terms of accuracy variance, the results illustrate that the variance is significantly greater for Regex and \ac{bf} models compared to \ac{rnn} and especially \ac{brnn} models, confirming observations from previous work on the initial dataset.
The near-perfect accuracy demonstrated by the proposed \ac{rnn} and \ac{brnn} models on the original dataset is robustly extended to the newly introduced and more extensive dataset. This showcases the models' effectiveness in maintaining exceptional accuracy even in the face of incomplete or incorrect language derivations across various programming languages and coverage tasks.

\begin{table*}[tb]
    \caption{Median values over \num{3} folds for the accuracy for snippets. The maximum scores per task are highlighted}
    \label{tab:results:accuracy_snippets}
    \centering
    \sisetup{table-format=1.4}
\rowcolors{2}{gray!10}{}
\begin{tabular}{
    l SSSS SSSS SSSS
}

\hiderowcolors
\toprule

\multirow{2}[2]{*}{\textbf{Model}} & \multicolumn{4}{c}{\textbf{\java}} & \multicolumn{4}{c}{\textbf{\kotlin}} & \multicolumn{4}{c}{\textbf{\python}} \\
\cmidrule(lr){2-5} \cmidrule(lr){6-9} \cmidrule(lr){10-13}
& {\textbf{\task{1}}} & {\textbf{\task{2}}} & {\textbf{\task{3}}} & {\textbf{\task{4}}} & {\textbf{\task{1}}} & {\textbf{\task{2}}} & {\textbf{\task{3}}} & {\textbf{\task{4}}} & {\textbf{\task{1}}} & {\textbf{\task{2}}} & {\textbf{\task{3}}} & {\textbf{\task{4}}} \\

\midrule
\showrowcolors

\apprbf{} & 0.9468 & 0.8420 & 0.7525 & 0.7356 & \tabhvalue 1.0000 & 0.9827 & 0.9765 & 0.9728 & \tabhvalue 1.0000 & \tabhvalue 1.0000 & \tabhvalue 1.0000 & \tabhvalue 1.0000 \\
\apprregex{} & 0.8856 & 0.7170 & 0.6653 & 0.6645 & 0.8447 & 0.6893 & 0.6570 & 0.6549 & 0.9401 & 0.8075 & 0.8071 & 0.8045 \\
\apprrnn{16} & \tabhvalue 1.0000 & 0.9579 & 0.9510 & 0.9505 & \tabhvalue 1.0000 & 0.9494 & 0.9460 & 0.9457 & \tabhvalue 1.0000 & 0.9603 & 0.9592 & 0.9586 \\
\apprrnn{32} & \tabhvalue 1.0000 & 0.9632 & 0.9555 & 0.9552 & \tabhvalue 1.0000 & 0.9528 & 0.9507 & 0.9504 & \tabhvalue 1.0000 & 0.9616 & 0.9613 & 0.9615 \\
\apprbrnn{16} & \tabhvalue 1.0000 & \tabhvalue 1.0000 & \tabhvalue 1.0000 & \tabhvalue 1.0000 & \tabhvalue 1.0000 & \tabhvalue 1.0000 & \tabhvalue 1.0000 & \tabhvalue 1.0000 & \tabhvalue 1.0000 & \tabhvalue 1.0000 & \tabhvalue 1.0000 & \tabhvalue 1.0000 \\
\apprbrnn{32} & \tabhvalue 1.0000 & \tabhvalue 1.0000 & \tabhvalue 1.0000 & \tabhvalue 1.0000 & \tabhvalue 1.0000 & \tabhvalue 1.0000 & \tabhvalue 1.0000 & \tabhvalue 1.0000 & \tabhvalue 1.0000 & \tabhvalue 1.0000 & \tabhvalue 1.0000 & \tabhvalue 1.0000 \\
\apprcnn{32} & \tabhvalue 1.0000 & \tabhvalue 1.0000 & \tabhvalue 1.0000 & \tabhvalue 1.0000 & \tabhvalue 1.0000 & \tabhvalue 1.0000 & \tabhvalue 1.0000 & \tabhvalue 1.0000 & \tabhvalue 1.0000 & \tabhvalue 1.0000 & \tabhvalue 1.0000 & \tabhvalue 1.0000 \\
\apprcnn{64} & \tabhvalue 1.0000 & \tabhvalue 1.0000 & \tabhvalue 1.0000 & \tabhvalue 1.0000 & \tabhvalue 1.0000 & \tabhvalue 1.0000 & \tabhvalue 1.0000 & \tabhvalue 1.0000 & \tabhvalue 1.0000 & \tabhvalue 1.0000 & \tabhvalue 1.0000 & \tabhvalue 1.0000 \\
\apprcnn{128} & \tabhvalue 1.0000 & \tabhvalue 1.0000 & \tabhvalue 1.0000 & \tabhvalue 1.0000 & \tabhvalue 1.0000 & \tabhvalue 1.0000 & \tabhvalue 1.0000 & \tabhvalue 1.0000 & \tabhvalue 1.0000 & \tabhvalue 1.0000 & \tabhvalue 1.0000 & \tabhvalue 1.0000 \setcounter{rownum}{0} \\

\hiderowcolors
\toprule

\textbf{Model} & \multicolumn{4}{c}{\textbf{\cpp}} & \multicolumn{4}{c}{\textbf{\csharp}} & \multicolumn{4}{c}{\textbf{\javascript}} \\

\midrule
\showrowcolors

\apprbf{} & \tabhvalue 1.0000 & \tabhvalue 1.0000 & \tabhvalue 1.0000 & \tabhvalue 1.0000 & \tabhvalue 1.0000 & \tabhvalue 1.0000 & \tabhvalue 1.0000 & \tabhvalue 1.0000 & \tabhvalue 1.0000 & 0.9784 & 0.9744 & 0.9744 \\
\apprregex{} & 0.9175 & 0.9722 & 0.8779 & 0.8779 & 0.8307 & 0.6841 & 0.6561 & 0.6561 & 0.9460 & 0.7949 & 0.7653 & 0.7653 \\
\apprrnn{16} & \tabhvalue 1.0000 & \tabhvalue 1.0000 & 0.9918 & 0.9959 & \tabhvalue 1.0000 & 0.9949 & 0.9891 & 0.9915 & \tabhvalue 1.0000 & 0.9193 & 0.9161 & 0.9227 \\
\apprrnn{32} & \tabhvalue 1.0000 & \tabhvalue 1.0000 & 0.9989 & 0.9957 & \tabhvalue 1.0000 & \tabhvalue 1.0000 & \tabhvalue 1.0000 & \tabhvalue 1.0000 & \tabhvalue 1.0000 & 0.8820 & 0.9284 & 0.9264 \\
\apprbrnn{16} & \tabhvalue 1.0000 & \tabhvalue 1.0000 & \tabhvalue 1.0000 & \tabhvalue 1.0000 & \tabhvalue 1.0000 & \tabhvalue 1.0000 & \tabhvalue 1.0000 & \tabhvalue 1.0000 & \tabhvalue 1.0000 & \tabhvalue 1.0000 & \tabhvalue 1.0000 & \tabhvalue 1.0000 \\
\apprbrnn{32} & \tabhvalue 1.0000 & \tabhvalue 1.0000 & \tabhvalue 1.0000 & \tabhvalue 1.0000 & \tabhvalue 1.0000 & \tabhvalue 1.0000 & \tabhvalue 1.0000 & \tabhvalue 1.0000 & \tabhvalue 1.0000 & \tabhvalue 1.0000 & \tabhvalue 1.0000 & \tabhvalue 1.0000 \\
\apprcnn{32} & \tabhvalue 1.0000 & \tabhvalue 1.0000 & \tabhvalue 1.0000 & \tabhvalue 1.0000 & \tabhvalue 1.0000 & \tabhvalue 1.0000 & \tabhvalue 1.0000 & \tabhvalue 1.0000 & \tabhvalue 1.0000 & \tabhvalue 1.0000 & \tabhvalue 1.0000 & \tabhvalue 1.0000 \\
\apprcnn{64} & \tabhvalue 1.0000 & \tabhvalue 1.0000 & \tabhvalue 1.0000 & \tabhvalue 1.0000 & \tabhvalue 1.0000 & \tabhvalue 1.0000 & \tabhvalue 1.0000 & \tabhvalue 1.0000 & \tabhvalue 1.0000 & \tabhvalue 1.0000 & \tabhvalue 1.0000 & \tabhvalue 1.0000 \\
\apprcnn{128} & \tabhvalue 1.0000 & \tabhvalue 1.0000 & \tabhvalue 1.0000 & \tabhvalue 1.0000 & \tabhvalue 1.0000 & \tabhvalue 1.0000 & \tabhvalue 1.0000 & \tabhvalue 1.0000 & \tabhvalue 1.0000 & \tabhvalue 1.0000 & \tabhvalue 1.0000 & \tabhvalue 1.0000 \\

\bottomrule

\end{tabular}

\end{table*}

\subsection{\req{4} -- \ac{cnn}: Accuracy}
\label{subsec:results:rq4}
\req{4} ascertains whether the \ac{cnn}32, \ac{cnn}64, and \ac{cnn}128 models can achieve the same near-perfect levels of \ac{sh} accuracy as the \emph{state-of-the-art} \ac{rnn}16, \ac{rnn}32, \ac{brnn}16, and \ac{brnn}32 models. The results are derived from the per-character \ac{sh} accuracy measured for each model concerning valid language derivations found in the extended \ac{sh} dataset. The results are summarised in Table \ref{tab:results:accuracy}.
The proposed \ac{cnn} models consistently deliver near-perfect \ac{sh} predictions across the five programming languages of \emph{Java}, \emph{Kotlin}, \emph{Python}, \emph{C++}, and \emph{JavaScript}. Only in the case of \emph{C\#} do the \ac{cnn} models exhibit a minor deviation from this trend, with a median accuracy rate in the high 99\%. Importantly, these near-perfect predictions remain consistent across all \emph{Coverage Tasks}.
Furthermore, the \ac{cnn} variants consistently outperform the non-bidirectional \ac{rnn}16 and \ac{rnn}32 models. These base \ac{rnn} models achieve comparable results in only specific tasks, such as \emph{T1} for \emph{Java}, \emph{Kotlin}, \emph{Python}, and \emph{JavaScript}, as well as \emph{T2} for \emph{C++}.
For each of the considered programming languages, the \ac{cnn}32, \ac{cnn}64, and \ac{cnn}128 models consistently produce \ac{sh} results that are tightly clustered around perfection, with only a minor number of outliers. This phenomenon is also observed in \emph{T4}, as depicted in Figure \ref{fig:results:boxplot_accuracy_t4}. The three \ac{cnn} variants do not introduce significant accuracy variations or trends and maintain a prediction density closely aligned with the \emph{state-of-the-art} bidirectional models \ac{brnn}16 and \ac{brnn}32.
Overall, the outcomes of \req{4} affirm that the proposed \ac{cnn} solutions do not result in observable losses in \ac{sh} accuracy. Instead, they demonstrate the potential to contribute on-par with the \emph{state-of-the-art} resolvers, thus providing a robust and viable alternative for \ac{sh}.

\subsection{\req{5} -- \ac{cnn}: Benchmarking}
\label{subsec:results:rq4}

\req{5} delves into the prediction speed of \ac{cnn} \ac{sh} models when executed on a \emph{CPU}, specifically examining their comparison to the instantaneous~\cite{10.1145/3540250.3549109, seow_designing_2008} response times observed for \ac{rnn} and \ac{brnn} models. This exploration serves the dual purpose of assessing the suitability of \ac{cnn}s running on \emph{CPU} for real-time applications, adhering to Seow's response-time categorisation, where instantaneous responses complete within 100 ms to 200 ms~\cite{seow_designing_2008}. Additionally, it seeks to identify potential speed-ups achievable through this execution approach.

Importantly, RQ7 will extend this inquiry to \ac{cnn} performance on \emph{GPU}, aligning with the proposed approach's intended usage and facilitating a comprehensive comparison against \emph{(B)RNN} variants.

The evaluation setup mirrors that of \req{2}, retraining and benchmarking \ac{cnn} models (\ac{cnn}32, \ac{cnn}64, and \ac{cnn}128) across \emph{Java}, \emph{Kotlin}, \emph{Python}, \emph{C++}, \emph{C\#}, and \emph{JavaScript}. The benchmarking is conducted on \emph{CPU} for \emph{T4}. The results are summarised in Table \ref{tab:results:time}.

For \emph{Python}, \ac{cnn}s emerge as the fastest models, surpassing \ac{rnn}s by 1.1 times and \ac{brnn}s by 1.3. Similarly, in \emph{JavaScript}, \ac{cnn}s perform the best, outpacing \ac{brnn}s by 1.7 times and \ac{rnn}s by 1.3.
In the case of \emph{Kotlin} and \emph{C\#}, \ac{cnn}s exhibit comparable performance to \ac{rnn}s while maintaining a 1.5 to 1.7 times speed advantage over \ac{brnn}s respectively. In the case of \emph{C++}, \ac{cnn}s perform on par with \ac{brnn}s, with \ac{rnn}s demonstrating a 2 times speed advantage.
In the case of \emph{Java}, \ac{cnn}s showcase a 1.7 times speed advantage over \ac{brnn}s, while \ac{rnn}s maintain a 1.2 times speed advantage over \ac{cnn}s.

Overall, when computed on \emph{CPU}s, \ac{cnn}s consistently achieve an instantaneous response time. They not only outperform \ac{brnn}s across various languages but also, in certain instances, compete favourably with the most lightweight alternatives: \ac{rnn} models. These findings position \ac{cnn}s as a promising choice for real-time \ac{sh} applications.

\subsection{\req{6} -- \ac{cnn}: Accuracy on Invalid Derivations}
\label{subsec:results:rq4}
\begin{figure*}[tb]
    \centering
    \includegraphics[width=1.0\linewidth]{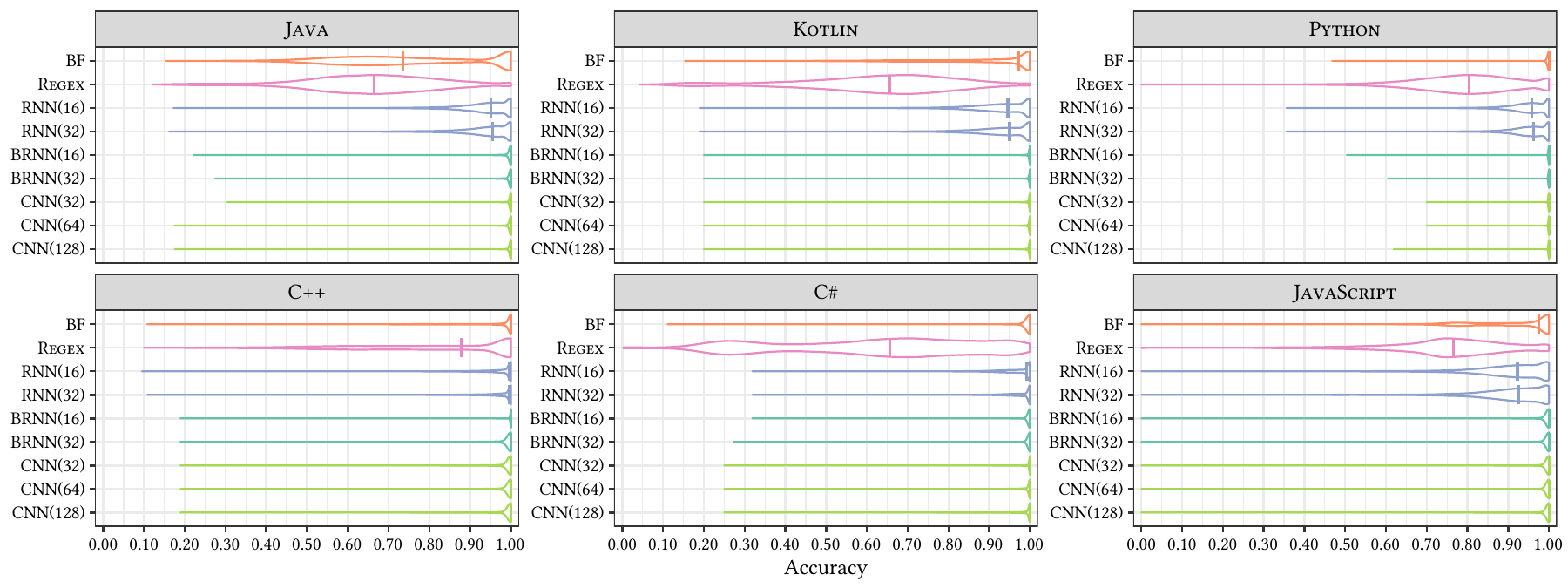}
    \caption{Accuracy values comparison for incomplete language derivations.}
    \label{fig:results:boxplot_accuracy_snippets}
\end{figure*}

To comprehensively evaluate the proposed \ac{cnn} models, \req{6} investigates the extent to which these models can maintain near-perfect accuracy in the face of incomplete or incorrect language derivations. Similar to \req{3}, the evaluation leverages the same snippet dataset, employing an evaluation strategy consistent with the \ac{rnn} models in the previous research question.
Table X reports that all \ac{cnn} models achieve near-perfect accuracy, comparable to the \ac{brnn}s. As illustrated in Figure \ref*{fig:results:boxplot_accuracy_snippets}, detailing the accuracy distribution for each model and language for \emph{T4}, \ac{cnn}32, \ac{cnn}64, and \ac{cnn}128 exhibit similar variance to \ac{brnn}32. The smaller \ac{cnn}32 model provides a slightly smaller variance advantage, particularly in \emph{Java} and \emph{Python}.
These findings affirm that the proposed \ac{cnn} models can indeed offer \ac{sh} of accuracy on par with the original \ac{cnn}-based approach, in the case of incorrect or incomplete language derivations.

\subsection{\req{7} -- \emph{GPU} Speed-Ups}
\label{subsec:results:rq4}

\req{7} delves into the examination of prediction speed-ups for deep models in the context of \ac{sh} when evaluated on a \emph{GPU}. Following the methodology akin to RQ2 and RQ5, all models trained for T4 undergo benchmarking on the same hardware for generating SH for all 20,000 files per language, repeating this process 30 times for statistical reliability.

The summarised results in Table \ref{tab:results:time}, denoted by the prefix ``G'' for \emph{GPU}, reveal that \emph{GPU} evaluation only yielded negligible improvements, with speed-ups reaching a maximum of 1.5 times faster for \emph{Java} and \emph{Kotlin}, followed by \emph{C\#} with 1.3, \emph{C++} with 1.2, \emph{Python} with 1.1, and \emph{JavaScript} performing on par. This aligns with the non-significant architectural optimisations observed for \ac{rnn}s when applied to \emph{GPU}s.
Similar to \ac{rnn}s, \emph{GPU} evaluation for \ac{brnn}s also produced marginal improvements, with 1.5 for \emph{Kotlin}, 1.3 for \emph{Java}, 1.2 for \emph{Python}, 1.1 for \emph{C++}, and \emph{C\#}, and performance on par for \emph{JavaScript}.

The intrinsically parallelisable nature of \ac{cnn} models resulted in significant speed-ups during \emph{GPU} evaluation. For \emph{Java}, \emph{GPU} evaluation led to 26.8 times faster predictions, 21.6 for \emph{C++}, 20.9 for \emph{C\#}, and 11.8 for \emph{Kotlin}. However, average improvements were negligible for languages with larger average sizes, such as \emph{Python} and \emph{JavaScript}, with improvements of 1.2 and 1.1, respectively. Despite this, the per-token performances of \emph{NN} models are equal for each language, and considering datasets with extremely large files, system integrators are anticipated to impose file size limits for code rendered in the browser.

In conclusion, while the architecture of \emph{(B)RNN} models resulted in limited improvements in prediction delays, the proposed approach relying on CNNs empowers GPUs to exploit their parallelisable architecture consistently, achieving the best prediction delays attainable for on-the-fly syntax highlighting.

\section{Related Work}
\label{sec:related_work}

The primary aim of the work presented here is to enhance the capabilities of real-time syntax highlighting tools by examining their generalisation abilities and offering improvements in evaluation speed. This research seeks to demonstrate the application of deep learning techniques to achieve not only effective but also efficient syntax highlighting. The subsequent section will outline the leading \emph{state-of-the-art} methodologies that bear the closest relevance to the approach being proposed, and how these differ.

\paragraph{Type Inference}
Deep learning has significantly influenced Type Inference, notably through \emph{DeepTyper}~\cite{hellendoorn_deep_2018}, \emph{Type4Py}~\cite{10.1145/3510003.3510124}, aiding the conversion from dynamically to statically typed languages. \emph{DeepTyper} employs a sophisticated bidirectional \emph{GRU}~\cite{cho_learning_2014} framework, introducing a distinctive Consistency Layer to improve handling of long-range inputs, and utilises a softmax function to assign type probabilities to each token. Contrary to the approach in this research, which concentrates on analysing sequences of token rules, \emph{DeepTyper} includes token identifiers to ascertain type names, a detail considered extraneous for the tasks at hand. \emph{Type4Py} advances this concept with a more complex neural network architecture and necessitates costly preprocessing steps like \ac{ast} derivation. However, the augmented complexity and processing requirements of models such as \emph{DeepTyper} and \emph{Type4Py} do not necessarily equate to enhanced performance for the methodology discussed here, highlighting a fundamental divergence in focus and efficacy between these models and the approach presented.
\textsc{TypeFix}, leveraging advanced transformer technology~\cite{vaswani_attention_2017}, serves as a decoder network for lenient parsing and typing of \emph{Java} code fragments, evolving from \textsc{DeepTyper}'s foundation~\cite{ahmed_learning_2021}. Its structure features a six-layer decoder, each layer enriched with multi-head attention and feed-forward mechanisms, enabling sophisticated handling of complex sequences. This design allows each layer's output to reflect a synthesis of all preceding unit combinations, facilitating the learning of generalizable input sequence patterns. The multi-headed attention further refines the model's capacity to discern intricate input relationships, surpassing traditional RNN models, which are limited by vanishing gradients~\cite{bengio_learning_1994}.
Like the \ac{sh} methodology and \textsc{DeepTyper}, \textsc{TypeFix} is trained using a synthetically created oracle, pairing \emph{Java} token identifiers with their deterministic types, thereby predicting categorical probability distributions across a defined type vocabulary.
Yet, this intricate architecture is not adopted for the immediate \ac{sh} approach, mirroring the considerations for \textsc{DeepTyper} and \textsc{Type4Py}. The emphasis remains on crafting more streamlined and effective models for real-time \ac{sh}, prioritising the unique demands and challenges of \ac{sh} over the complexity offered by \textsc{TypeFix}.

\paragraph{Island Grammars}
Island Grammars introduce a framework for grammar design, segregating grammar rules into ``island'' for specific subsequences and ``water'' for the remaining tokens~\cite{moonen_generating_2001}. This structure allows for targeted processing of sequences relevant for highlighting in \ac{sh} tasks, with ``island'' rules focusing on highlight-worthy sequences and ``water'' rules managing the rest.
Despite its potential, this methodology diverges from the current research's trajectory~\cite{10.1145/3540250.3549109}. Crafting an island grammar demands a deep understanding of grammatical constructs and a meticulous definition process for productions, a task more intricate than creating a tree walker for existing grammars. This complexity contrasts with the current research's goal of simplifying development and improving automation in SH tasks.
Moreover, island grammars do not fully resolve the challenge of processing incomplete language derivations, a limitation shared with the state-of-practice approach that this research intends to transcend. Hence, island grammars do not meet the aims of this research, which prioritises more streamlined and automated strategies for SH tasks.

\paragraph{Program Synthesis}
The process described in this line of work also diverges fundamentally from the of Program Synthesis, which is concerned with the generation of programs that map inputs to outputs. Program Synthesis, exemplified by projects like \emph{DeepCoder}~\cite{deepcoder} and \emph{PQT}~\cite{abolafia2018neural}, aims to infer program structures to bind inputs with outputs, enhancing traditional search techniques through predictive neural networks. These models complement rather than replace search-based methods, focusing on program generation guided by input-output examples rather than understanding the intricacies of compilers or interpreters.
Techniques such as execution-guided synthesis in \emph{Execution-Guided Neural Program Synthesis}~\cite{chen2018execution} aim to improve predictions based on program state manipulations, distinct from the operational semantics learning associated with compilers. Similar approaches enhance the synthesis or search process and rule generation, steering clear of mimicking compiler or interpreter functionalities ~\cite{alford2022neural, noriega2022neural}.
\emph{NGST2}~\cite{mariano2022automated} introduces a formal method for program translation through trace-compatibility and cognate grammars, focusing on syntactic conversion between programming paradigms rather than delving into the mechanics of program execution akin to a compiler or interpreter. This highlights a clear distinction from the process presented in this work, which seeks to understand and replicate the mapping of inputs to outputs in the manner of compilers or interpreters, setting it apart from the objectives and methodologies of Program Synthesis.

\section{Conclusions and Future Work}
\label{sec:conclusions}



This work advanced the domain of \emph{On-the-Fly Syntax Highlighting}. It delivered an extended dataset to include six programming languages. Now including \emph{C++}, \emph{C\#}, and \emph{JavaScript}, in addition to the original \emph{Java}, \emph{Kotlin}, and \emph{Python}, this expansion not only enriches benchmarking capabilities but also broadens the scope of application in diverse coding environments. The investigation into the generalisation capabilities of \emph{state-of-the-art} \ac{rnn} and \ac{brnn} methodologies has yielded promising results, demonstrating robustness with near-perfect accuracy and manageable time delays. 
The precision and efficiency demonstrated in benchmarking evaluations indicate that the \ac{cnn} method not only maintains near-perfect predictions but does so at a significantly faster rate, especially when evaluated on \emph{GPU} platforms. This positions the \emph{CNN}-based approach as the front-runner in the realm of \emph{On-the-Fly Syntax Highlighting}, both in terms of accuracy and speed.

Future work should also consider the efficiency of the training process. Preliminary investigations suggest that there is potential for reducing the number of training samples, which could lead to a significant decrease in training costs. Furthermore, the exploration of multilingual models would be a logical extension, potentially streamlining deployment in diverse programming environments and thereby increasing practical applicability.
It is important to recognise that applying the principles of this research in other fields will inherently lead to improvements in both the development processes and the tooling. Employing this technology to recognise and interpret code snippets from diverse web sources, even those not strictly adhering to standard syntax, may boost how developers engage with code on various platforms. This will not only enhance tools for code comprehension and error detection but also refine the processes involved in developing these tools. Developers could shift their focus to creating straightforward, brute-force solutions, focusing less on performance optimisation or tolerance to noisy inputs. Simultaneously, the tools themselves are set to be more accurate and responsive. This dual advancement in both process and tooling promises a transformative impact on the software development lifecycle.


\section*{Acknowledgements}

The research leading to these results has received funding from the Swiss National Science Foundation (SNSF) project \enquote{Melise - Machine Learning Assisted Software Development} (SNSF204632).

\balance
\bibliography{references, urls}

\begin{thebibliography}{10}
\providecommand{\url}[1]{#1}
\csname url@samestyle\endcsname
\providecommand{\newblock}{\relax}
\providecommand{\bibinfo}[2]{#2}
\providecommand{\BIBentrySTDinterwordspacing}{\spaceskip=0pt\relax}
\providecommand{\BIBentryALTinterwordstretchfactor}{4}
\providecommand{\BIBentryALTinterwordspacing}{\spaceskip=\fontdimen2\font plus
\BIBentryALTinterwordstretchfactor\fontdimen3\font minus \fontdimen4\font\relax}
\providecommand{\BIBforeignlanguage}[2]{{%
\expandafter\ifx\csname l@#1\endcsname\relax
\typeout{** WARNING: IEEEtran.bst: No hyphenation pattern has been}%
\typeout{** loaded for the language `#1'. Using the pattern for}%
\typeout{** the default language instead.}%
\else
\language=\csname l@#1\endcsname
\fi
#2}}
\providecommand{\BIBdecl}{\relax}
\BIBdecl

\bibitem{sarkar_impact_2015}
A.~Sarkar, ``The {{Impact}} of {{Syntax Colouring}} on {{Program Comprehension}},'' in \emph{Annual {{Meeting}} of the {{Psychology}} of {{Programming Interest Group}} ({{PPIG}})}, 2015.

\bibitem{10.1145/3540250.3549109}
\BIBentryALTinterwordspacing
M.~E. Palma, P.~Salza, and H.~C. Gall, ``On-the-fly syntax highlighting using neural networks,'' in \emph{Proceedings of the 30th ACM Joint European Software Engineering Conference and Symposium on the Foundations of Software Engineering}, ser. ESEC/FSE 2022.\hskip 1em plus 0.5em minus 0.4em\relax New York, NY, USA: Association for Computing Machinery, 2022, p. 269–280. [Online]. Available: \url{https://doi.org/10.1145/3540250.3549109}
\BIBentrySTDinterwordspacing

\bibitem{replicationpackage}
\BIBentryALTinterwordspacing
M.~E. Palma, A.~Wolf, P.~Salza, and H.~C. Gall. (2024) {On-the-Fly Syntax Highlighting Generalisability and Speed-ups - Replication Package}. [Online]. Available: \url{https://doi.org/10.5281/zenodo.10655088}
\BIBentrySTDinterwordspacing

\bibitem{sutskever_sequence_2014}
I.~Sutskever, O.~Vinyals, and Q.~V. Le, ``Sequence to {{Sequence Learning}} with {{Neural Networks}},'' in \emph{International {{Conference}} on {{Neural Information Processing Systems}} ({{NIPS}})}, 2014, pp. 3104--3112.

\bibitem{cho_learning_2014}
K.~Cho, B.~{van Merrienboer}, C.~Gulcehre, D.~Bahdanau, F.~Bougares, H.~Schwenk, and Y.~Bengio, ``Learning {{Phrase Representations Using RNN Encoder}}\textendash{{Decoder}} for {{Statistical Machine Translation}},'' in \emph{Conference on {{Empirical Methods}} in {{Natural Language Processing}} ({{EMNLP}})}, 2014, pp. 1724--1734.

\bibitem{schuster_bidirectional_1997}
M.~Schuster and K.~K. Paliwal, ``Bidirectional {{Recurrent Neural Networks}},'' \emph{Ieee Transactions on Signal Processing}, vol.~45, no.~11, pp. 2673--2681, 1997.

\bibitem{gehring2017convolutional}
J.~Gehring, M.~Auli, D.~Grangier, D.~Yarats, and Y.~N. Dauphin, ``Convolutional sequence to sequence learning,'' in \emph{International conference on machine learning}, 2017, pp. 1243--1252.

\bibitem{nguyen_2016}
N.~Ngoc~Giang, V.~Tran, D.~Ngo, D.~Phan, F.~Lumbanraja, M.~R. Faisal, B.~Abapihi, M.~Kubo, and K.~Satou, ``Dna sequence classification by convolutional neural network,'' \emph{Journal of Biomedical Science and Engineering}, vol.~09, pp. 280--286, 01 2016.

\bibitem{antlr}
\BIBentryALTinterwordspacing
T.~Parrm. (2022) {ANTLR}. [Online]. Available: \url{https://www.antlr.org}
\BIBentrySTDinterwordspacing

\bibitem{pygments}
\BIBentryALTinterwordspacing
G.~Brandl. (2022) {Pygments}. [Online]. Available: \url{https://pygments.org}
\BIBentrySTDinterwordspacing

\bibitem{abadi_tensorflow_2015}
\BIBentryALTinterwordspacing
M.~Abadi, A.~Agarwal, P.~Barham, E.~Brevdo, Z.~Chen, C.~Citro, G.~S. Corrado, A.~Davis, J.~Dean, M.~Devin, S.~Ghemawat, I.~Goodfellow, A.~Harp, G.~Irving, M.~Isard, Y.~Jia, R.~Jozefowicz, L.~Kaiser, M.~Kudlur, J.~Levenberg, D.~Man{\'e}, R.~Monga, S.~Moore, D.~Murray, C.~Olah, M.~Schuster, J.~Shlens, B.~Steiner, I.~Sutskever, K.~Talwar, P.~Tucker, V.~Vanhoucke, V.~Vasudevan, F.~Vi{\'e}gas, O.~Vinyals, P.~Warden, M.~Wattenberg, M.~Wicke, Y.~Yu, and X.~Zheng, ``{{TensorFlow}}: {{Large-Scale Machine Learning}} on {{Heterogeneous Systems}},'' 2015. [Online]. Available: \url{https://www.tensorflow.org}
\BIBentrySTDinterwordspacing

\bibitem{montgomery_design_2017}
D.~C. Montgomery, \emph{Design and {{Analysis}} of {{Experiments}}}.\hskip 1em plus 0.5em minus 0.4em\relax {Wiley}, 2017.

\bibitem{vargha_critique_2000}
A.~Vargha and H.~D. Delaney, ``A {{Critique}} and {{Improvement}} of the "{{CL}}" {{Common Language Effect Size Statistics}} of {{McGraw}} and {{Wong}},'' \emph{Journal of Educational and Behavioral Statistics}, vol.~25, no.~2, pp. 101--132, 2000.

\bibitem{seow_designing_2008}
S.~C. Seow, \emph{Designing and {{Engineering Time}}: {{The Psychology}} of {{Time Perception}} in {{Software}}}.\hskip 1em plus 0.5em minus 0.4em\relax {Addison-Wesley Professional}, 2008.

\bibitem{dabrowski_40_2011}
J.~Dabrowski and E.~V. Munson, ``40 {{Years}} of {{Searching}} for the {{Best Computer System Response Time}},'' \emph{Interacting with Computers}, vol.~23, no.~5, pp. 555--564, 2011.

\bibitem{hellendoorn_deep_2018}
V.~J. Hellendoorn, C.~Bird, E.~T. Barr, and M.~Allamanis, ``Deep {{Learning Type Inference}},'' in \emph{{{ACM Joint European Software Engineering Conference}} and {{Symposium}} on the {{Foundations}} of {{Software Engineering}} ({{ESEC}}/{{FSE}})}, 2018, pp. 152--162.

\bibitem{10.1145/3510003.3510124}
\BIBentryALTinterwordspacing
A.~M. Mir, E.~Lato\v{s}kinas, S.~Proksch, and G.~Gousios, ``Type4py: Practical deep similarity learning-based type inference for python,'' in \emph{Proceedings of the 44th International Conference on Software Engineering}, ser. ICSE '22.\hskip 1em plus 0.5em minus 0.4em\relax New York, NY, USA: Association for Computing Machinery, 2022, pp. 2241--2252. [Online]. Available: \url{https://doi.org/10.1145/3510003.3510124}
\BIBentrySTDinterwordspacing

\bibitem{vaswani_attention_2017}
A.~Vaswani, N.~Shazeer, N.~Parmar, J.~Uszkoreit, L.~Jones, A.~N. Gomez, {\L}.~Kaiser, and I.~Polosukhin, ``Attention {{Is All You Need}},'' in \emph{Conference on {{Neural Information Processing Systems}} ({{NIPS}})}, I.~Guyon, U.~V. Luxburg, S.~Bengio, H.~Wallach, R.~Fergus, S.~Vishwanathan, and R.~Garnett, Eds., 2017, pp. 5998--6008.

\bibitem{ahmed_learning_2021}
T.~Ahmed, P.~Devanbu, and V.~J. Hellendoorn, ``Learning {{Lenient Parsing}} \& {{Typing Via Indirect Supervision}},'' \emph{Empirical Software Engineering}, vol.~26, no.~2, pp. 1--31, 2021.

\bibitem{bengio_learning_1994}
Y.~Bengio, P.~Simard, and P.~Frasconi, ``Learning {{Long-Term Dependencies}} with {{Gradient Descent Is Difficult}},'' \emph{Ieee Transactions on Neural Networks}, vol.~5, no.~2, pp. 157--166, 1994.

\bibitem{moonen_generating_2001}
L.~Moonen, ``Generating {{Robust Parsers Using Island Grammars}},'' in \emph{Working {{Conference}} on {{Reverse Engineering}} ({{WCRE}})}, 2001, pp. 13--22.

\bibitem{deepcoder}
M.~Balog, A.~L. Gaunt, M.~Brockschmidt, S.~Nowozin, and D.~Tarlow, ``Deepcoder: Learning to write programs,'' \emph{arXiv preprint arXiv:1611.01989}, 2016.

\bibitem{abolafia2018neural}
D.~A. Abolafia, M.~Norouzi, J.~Shen, R.~Zhao, and Q.~V. Le, ``Neural program synthesis with priority queue training,'' \emph{arXiv preprint arXiv:1801.03526}, 2018.

\bibitem{chen2018execution}
X.~Chen, C.~Liu, and D.~Song, ``Execution-guided neural program synthesis,'' in \emph{International Conference on Learning Representations}, 2018.

\bibitem{alford2022neural}
S.~Alford, A.~Gandhi, A.~Rangamani, A.~Banburski, T.~Wang, S.~Dandekar, J.~Chin, T.~Poggio, and P.~Chin, ``Neural-guided, bidirectional program search for abstraction and reasoning,'' in \emph{Complex Networks \& Their Applications X: Volume 1, Proceedings of the Tenth International Conference on Complex Networks and Their Applications COMPLEX NETWORKS 2021 10}.\hskip 1em plus 0.5em minus 0.4em\relax Springer, 2022, pp. 657--668.

\bibitem{noriega2022neural}
E.~Noriega-Atala, R.~Vacareanu, G.~Hahn-Powell, and M.~A. Valenzuela-Esc{\'a}rcega, ``Neural-guided program synthesis of information extraction rules using self-supervision,'' in \emph{Proceedings of the First Workshop on Pattern-based Approaches to NLP in the Age of Deep Learning}, 2022, pp. 85--93.

\bibitem{mariano2022automated}
B.~Mariano, Y.~Chen, Y.~Feng, G.~Durrett, and I.~Dillig, ``Automated transpilation of imperative to functional code using neural-guided program synthesis,'' \emph{Proceedings of the ACM on Programming Languages}, vol.~6, no. OOPSLA1, pp. 1--27, 2022.

\end{thebibliography}

\begin{complete-version}
\begin{IEEEbiography}%
	[{\includegraphics[width=1in,height=1.25in,clip,keepaspectratio]{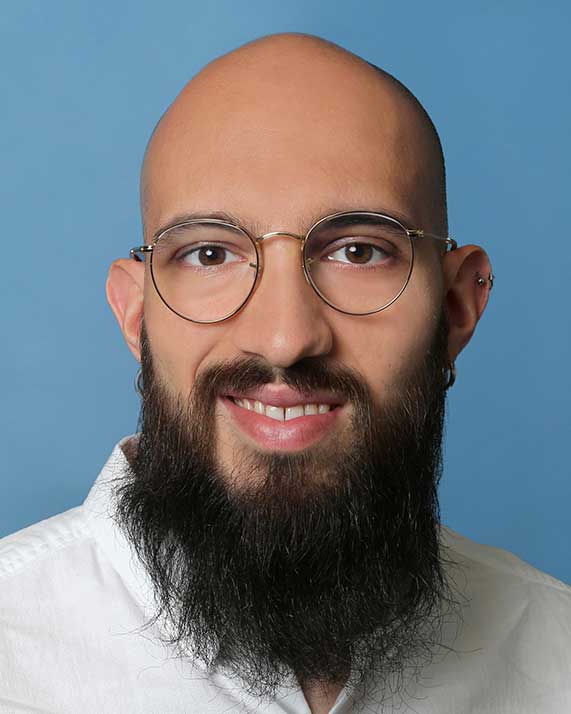}}]
	{Pasquale Salza}%
	is a Senior Research Associate
    in the Software Evolution and Architecture Lab (s.e.a.l.)
    at the University of Zurich, Switzerland.
    He received a Ph.D. degree in Computer Science from the University of Salerno, Italy.
	His research interests include software engineering, machine learning, cloud computing, and evolutionary computation.
	Contact him at \href{mailto:salza@ifi.uzh.ch}{salza@ifi.uzh.ch}.
\end{IEEEbiography}

\begin{IEEEbiography}%
	[{\includegraphics[width=1in,height=1.25in,clip,keepaspectratio]{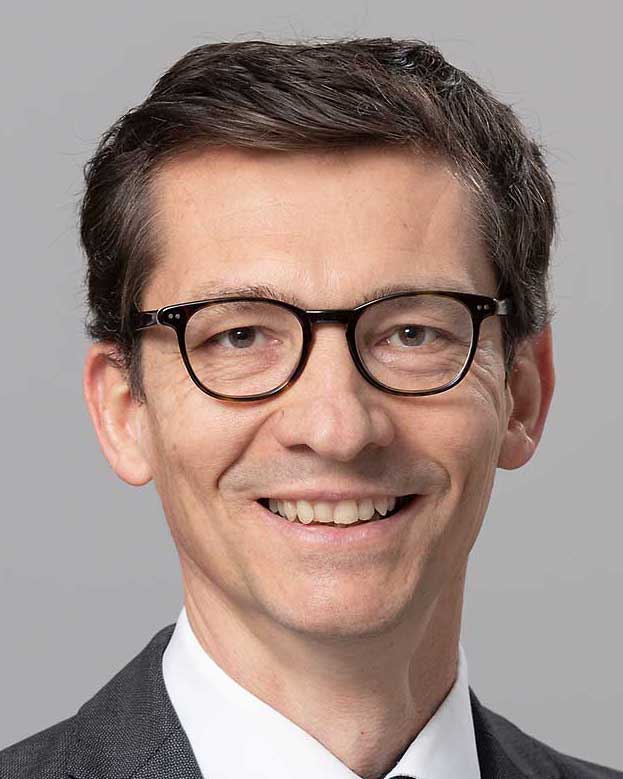}}]
    {Harald C. Gall}%
    is Dean of the Faculty of Business, Economics, and Informatics at the University of Zurich.
    He is professor of software engineering in the Department of Informatics.
    He held visiting positions at Microsoft Research in Redmond, USA, and University of Washington in Seattle, USA.
    His research interests are software evolution, software architecture, software quality, and cloud-based software engineering.
    Since 1997, he has worked on devising ways in which mining repositories can help to better understand and improve software development.
    Contact him at \href{mailto:gall@ifi.uzh.ch}{gall@ifi.uzh.ch}.
\end{IEEEbiography}

\vfill

\end{complete-version}

\end{document}